\documentclass{aa}
\usepackage{graphicx}
\usepackage{natbib}
\usepackage{txfonts}
\begin{document}
   \title{The Modelling of InfraRed Dark Cloud Cores}
   \author{C.W.~Ormel
          \inst{1}
          \and
          R.F.~Shipman
          \inst{1,2}
          \and
          V.~Ossenkopf
          \inst{2,3}
          \and
          F.P.~Helmich
          \inst{1,2}
          }
   \offprints{ormel@astro.rug.nl}
   \institute{Kapteyn Astronomical Institute, University of Groningen, PO box 800, 9700 AV  Groningen, The Netherlands\\
             \email{ormel@astro.rug.nl}
             \and
              SRON National Institute for Space Research, Landleven 12, 9747 AD Groningen, The Netherlands\\
             \email{russ@sron.rug.nl}\\
             \email{f.p.helmich@sron.rug.nl}
             \and
             {\sc i.} Physikalisches Institut, Universit\"at zu K\"oln, Z\"ulpicher Stra\ss e 77, 50937 K\"oln, Germany\\
             \email{ossk@ph1.uni-koeln.de}
            }
   \date{}
             
   \abstract{This paper presents results from modelling $450\ \mu\mathrm{m}$ and $850\ \mu\mathrm{m}$ continuum and \element[+]{HCO} line observations of three distinct cores of an infrared dark cloud (IRDC) directed toward the \object{W51} GMC. In the sub-mm continuum these cores appear as bright, isolated emission features. One of them coincides with the peak of $8.3\ \mu\mathrm{m}$ extinction as measured by the Midcourse Space Experiment satellite. Detailed radiative transfer codes are applied to constrain the cores' physical conditions to address the key question: Do these IRDC-cores harbour luminous sources? The results of the continuum model, expressed in the $\chi^2$ quality-of-fit parameter, are also constrained by the absence of $100\ \mu\mathrm{m}$ emission from IRAS. For the sub-mm emission peaks this shows that sources of $\sim 300\ \mathrm{L}_{\sun}$ are embedded within the cores. For the extinction peak, the combination of continuum and \element[+]{HCO} line modelling indicates that a heating source is present as well. Furthermore, the line model provides constraints on the clumpiness of the medium. All three cores have similar masses of about $70-150\ \mathrm{M}_{\sun}$ and similar density structures. The extinction peak differs from the other two cores by hosting a much weaker heating source, and the sub-mm emission core at the edge of the IRDC deviates from the other cores by a higher internal clumpiness. 
   
   }

   \maketitle
\section{Introduction}
InfraRed Dark Clouds (IRDCs) have been observed in recent mid-IR surveys of the Galactic Plane \citep{1996AA...315L.165P, 1998ApJ...494L.199E} as prominent absorption features against the bright mid-IR Galactic emission. Due to lack of $60\ \mu\mathrm{m}$ and  $100\ \mu\textrm{m}$ emission, IRDCs must be cold, i.e., below $13\ \mathrm{K}$ \citep{1998ApJ...494L.199E}. This cold nature was confirmed, by follow-up observations of molecular lines and sub-mm continuum observations and placed most IRDCs below $15\ \mathrm{K}$ \citep{1998ApJ...508..721C,2002AA...382..624T}. Furthermore, these line observations provided radial velocities and hence estimates of the distances. Most IRDCs are at distances above $1\ \textrm{kpc}$ and are preferentially located toward spiral arms and the Galactic Ring \citep{1999usis.conf..671E, 2000AAS...197.0516C}.

Thousands of IRDCs have been found throughout the Galactic Plane and counterparts are found at different wavelengths \citep{2001AAS...199.9706S}. \citet{2000ApJ...543L.157C} identified a number of IRDCs from the Midcourse Space eXperiment (MSX) Galactic Plane survey \citep{2001AJ....121.2819P} and observed these IRDCs in $850\ \mu\mathrm{m}$ dust continuum emission. They found a variety of sources or cores associated with the IRDCs. Most of these `sources' are seen in absorption at $8.3\ \mu\textrm{m}$, but occasionally also in emission. \citet{2000ApJ...543L.157C} found masses ranging from $20\ \textrm{M}_{\sun}$ to $1000\ \mathrm{M}_{\sun}$ and temperatures from cold ($10\ \mathrm{K}$) to warm ($40-50\ \mathrm{K}$). \citet{2003ApJ...586.1127R}, using SCUBA, examined the \object{G79.3+0.3} region and found young stellar objects to be associated with sub-mm observations at the edge of the IRDC complex.

Because of their sheer sizes, high column densities and low temperatures, it has been suggested that IRDCs are the sites of the formation of massive stars. However, unlike low-mass star formation, little observational and theoretical information is available on the earliest stages of the formation of high-mass stars. Many high-mass star formation sites are known, but they are always in a relatively evolved state, where temperatures are $> 30\ \mathrm{K}$ rather than $10\ \mathrm{K}$. Therefore, if high-mass star formation can be linked to IRDCs, this would complement the evolutionary picture.

Low-mass cores have been studied in great detail in the sub-mm. Models of the spectral energy distribution allow to constrain the density and temperature structures of the cores to be compared with theoretical models. These observations generally support theories of star formation in molecular cores \citep{2000ApJS..131..249S, 2001ApJ...557..193E, 2002ApJ...575..337S}. Similar techniques have been applied to high-mass protostars in order to probe the physical structure of the envelopes surrounding these sources \citep{2000ApJ...537..283V, 2003AA...409..589H, 2002ApJS..143..469M}. The difference between the observations described here and those listed above, is that all but two of the sources listed in the literature have strong mid- to far-IR emission which indicates the sources are relatively well evolved. The two sources of \citet{2002ApJS..143..469M}, which show no mid-IR emission, were only observed at $350\ \mu\mathrm{m}$ and therefore not modelled.  

There are many questions regarding IRDCs. Do they represent the earliest stages of high-mass star formation? If so, how do they compare with Bok globules and low-mass star formation sites? If not, what is their fate? In order to begin addressing these questions, SCUBA $450\ \mu\textrm{m}$ and $850\ \mu\textrm{m}$ data have been obtained toward one IRDC complex seen in the direction of the \object{W51} GMC. Furthermore, \element[][]{H}\element[+][]{CO} line observations were taken. This article presents and discusses the physical properties (density and temperature structure and internal heating) of the complex through detailed modelling of both the sub-mm dust continuum emission and the \element[+]{HCO} line data.  
\begin{figure}
    \centering
    \includegraphics[width=88mm, bb = 20 20 410 430, clip = true]{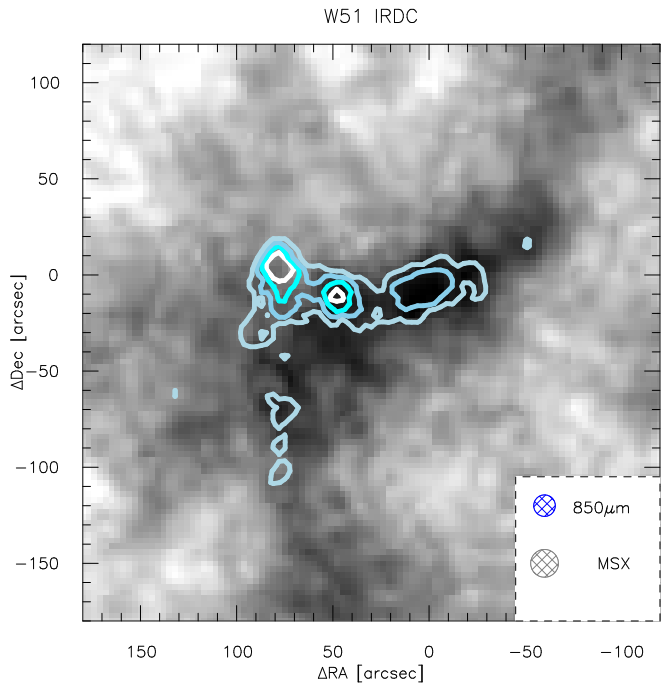}
    \caption{InfraRed Dark Cloud in the direction of \object{W51} as observed by MSX with SCUBA contours overlaid. The MSX image is stretched from \mbox{$2.0 \times 10^{-6}\ \mathrm{W\ m}^{-2}\ \mathrm{sr}^{-1}$} (black) to \mbox{$4.5\times 10^{-6}\ \mathrm{W\ m}^{-2}\ \mathrm{sr}^{-1}$} (white). Overlaid are SCUBA emission contours at 4,6,8 and 10 times the  $850\ \mu\mathrm{m}$ rms noise level of $\sigma_{850} = 0.040\ \mathrm{Jy\ beam}^{-1}$. The zero point of this image is at 19$^h$21$^m$44.7$^s$, 13$^\circ$49$'$34.7$''$ (J2000).}
    \label{fig:W51}
\end{figure}

Section \ref{sec:obs} discusses the observational data and the data reduction. In Sect. \ref{sec:mirext} the SCUBA and MSX observations are compared to estimate the optical depth. In Sect. \ref{sec:cont} the continuum data is modelled using a spherically symmetric core model with the aim to clarify the density and temperature structures. These physical structures are used as input for the modelling of the \element[+]{HCO} line data (Sect. \ref{sec:line}), from which constraints on the velocity structure and the clumpiness are derived. Finally, an overall discussion is presented in Sect. \ref{sec:discuss}.

\section{Observations\label{sec:obs}}
\subsection{MSX Data}
This investigation has made use of MSX Galactic Plane survey data \citep{2001AJ....121.2819P}. The MSX radiometer, SPIRIT III, has surveyed the Galactic Plane in four mid-IR bands with 30 times higher spatial resolution than IRAS.  The MSX spectral bands are centred at 8.3, 12.1, 14.7 and 21.3 $\mu\mathrm{m}$. The $8.3\ \mu\mathrm{m}$  spectral band is especially sensitive and covers the brightest of the Galactic PAH features between $6-9\ \mu\mathrm{m}$. It is against this Galactic emission that IRDCs are seen. 

The IRDC considered in this article is a complex clearly visible in MSX images  in the direction of the \object{W51} GMC (Fig. \ref{fig:W51}). The image is in equatorial coordinates and has been centred on the peak extinction  which is also at the centre of the S-shaped extinction valley. The greyscale is chosen such that black regions correspond to areas of lowest $8.3\ \mu\mathrm{m}$ emission, or highest extinction by the IRDC. Section \ref{sec:mirext} elaborates on the MSX extinction and on its relation to the SCUBA $850\ \mu\mathrm{m}$ emission, which is shown by the overlaid countours in Fig. \ref{fig:W51}.

\subsection{JCMT Continuum Observations}
JCMT\footnote{The JCMT is operated by the Joint Astronomy Centre in Hilo, Hawaii on behalf of the parent organisations Particle Physics and Astronomy Research Council in the United Kingdom, the National Research Council of Canada and The Netherlands Organisation for Scientific Research.} observations of the \object{W51-IRDC} were made on October 28, 2002 at $450\ \mu\mathrm{m}$ and $850\ \mu\mathrm{m}$ with the Sub-millimeter Common User Bolometer Array. The SCUBA observations had good weather conditions with an atmospheric optical depth at $225\ \mathrm{GHz}$ of $\tau_\mathrm{CSO} \sim 0.06$.  To cover the extinction region seen in MSX, a 6$^\prime$ by 4$^\prime$ area was observed using the scan map mode. Two perpendicular chop angles were used with three chop throws of 30$\arcsec$, 44$\arcsec$, 68$\arcsec$ and the image was reconstructed using the Emerson II technique \citep{2000adass...9..559J}. With the Emerson II image reconstruction technique, spatial structures much longer than the 68\arcsec chop throw are not reliably reconstructed \citep{2003ApJ...588L..37J}. This implies that a zero level correction is necessary and that this correction is local, i.e., it changes over the image on arcminute scales.

The data were flat-fielded, extinction corrected, and calibrated on Uranus using the standard SCUBA reduction software \citep{1998asp...145..216J}.  During the observations, SCUBA was experiencing the transient noise phenomenon associated with contamination caused by superfluid helium films.  Over a few days, this film moved along the bolometer arrays, and in this case, moved from the $850\ \mu\textrm{m}$ array to the $450\ \mu\textrm{m}$ array. However, over the period of a few hours the noise was stable enough to identify affected bolometers and to remove them from further processing.  The removal of these bolometers had as main result the increase of the noise in the final image above the value expected for the assigned observing time.  From the final images the noise was estimated to be $250\ \mathrm{mJy\ beam^{-1}}$ at $450\ \mu\mathrm{m}$ and $40\ \mathrm{mJy\ beam^{-1}}$ at $850\ \mu\mathrm{m}$.
\subsubsection{Calibration and Image Reconstruction}
Although the calibration should be good to 15\% and 25\% for the $850\ \textrm{W}$ and $450\ \textrm{W}$ filters, respectively \citep{2000adass...9..559J}, the observations were obtained early in the evening when the Flux Conversion Factors (FCF: the conversion from engineering units of Volts to $\mathrm{Jy\ \mathrm{beam}^{-1}}$) can change significantly during the time of the observations \citetext{Wouterloot, priv.\ comm.}. For the observations presented in this article the FCFs are likely to have declined by about 10\% and 20\%, respectively, for $850\ \mu\mathrm{m}$ and $450\ \mu\mathrm{m}$. Unfortunately, Uranus observations were not obtained before \emph{and} after the IRDC observation, thus a potential systematic calibration error had to be taken into account.

It is, however, possible to estimate the impact of this calibration change by re-performing the fit of Sect. \ref{sec:results} with the artificial reduction of the $850\ \mu\mathrm{m}$ and  $450\ \mu\mathrm{m}$ calibration factors by 10\% and 20\%, respectively. The main effect of this reduction is to broaden the $\chi^2$-topology, while the best fitting parameters are not changed significantly. A varying FCF, although increasing the uncertainty in the calibration, does not alter the main findings of this work. 

\subsubsection{SCUBA Beams}
\begin{figure}[tp]
    \centering
    \includegraphics[width=43mm]{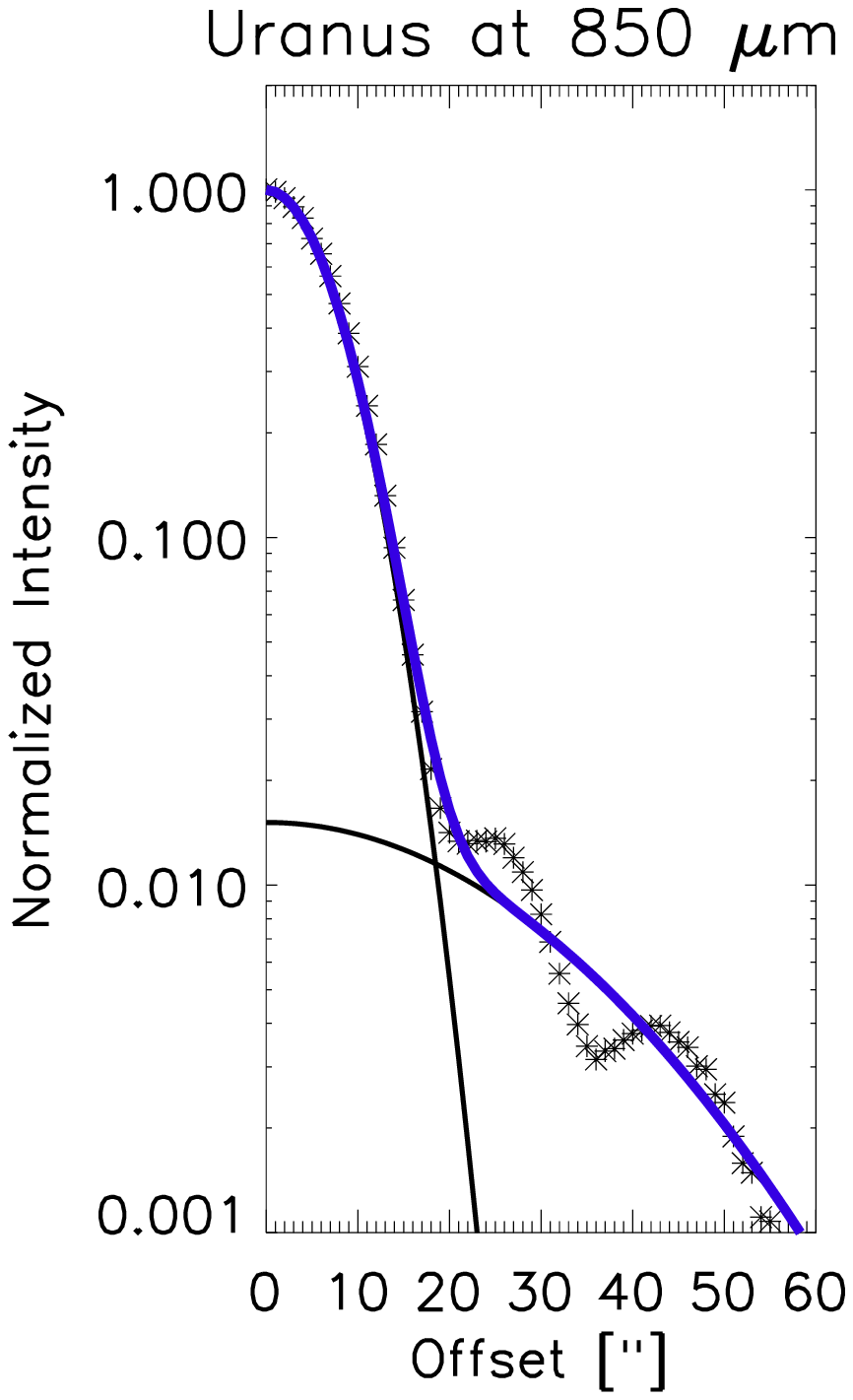}
    \includegraphics[width=43mm]{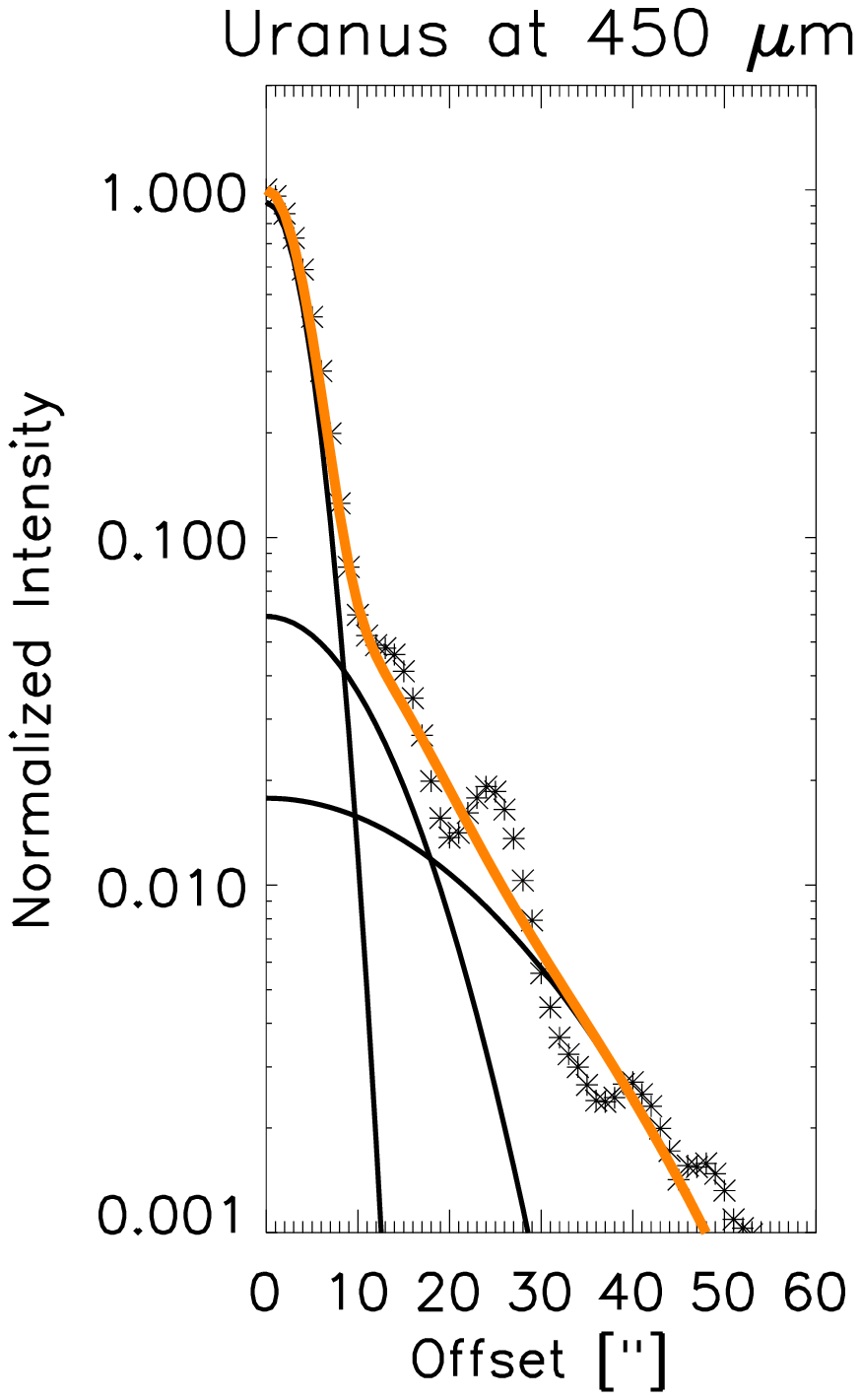}
    \caption{Beam profiles at $850\ \mu\mathrm{m}$ (left) and $450\ \mu\mathrm{m}$ (right) as measured on Uranus on October 28, 2002. The beam profile is represented by a superposition of Gaussians, whose parameters are listed in Table \ref{tab:beam}. The beam profiles are normalised to their central ($\theta = 0''$) value.}
    \label{fig:beams}
\end{figure}

\begin{table}
    \begin{center}
    \begin{tabular}{lll}
    \multicolumn{3}{c}{\sc SCUBA beam pattern}\\
    \hline
    \hline
    $\lambda$   & relative power $w_i$ &   FWHM    \\
    $[\mu\mathrm{m}]$ &             &   [$''$]\\
    \hline      
    450\dots\dots& 0.45             &   8.0 \\
                & 0.25              &  23.5\\
                & 0.30              &  47.0\\
    850\dots\dots& 0.80             &  14.6\\
                & 0.20              &  58.8\\
    \hline
    \end{tabular}
    \caption{Fit to the SCUBA beam pattern as observed on Uranus. The total beam is approximated to be a superposition of Gaussians. The $450\ \mu\textrm{m}$ beam strongly deviates from a single Gaussian, with more than 50\% of the power falling into the error beams. The beam profiles are plotted in Fig. \ref{fig:beams}.}
    \label{tab:beam}
    \end{center}
\end{table}
Since both SCUBA $450\ \mu\textrm{m}$ and $850\ \mu\textrm{m}$ beam patterns are quite complex and the modelling was restricted to one dimension, azimuthal averages of the Uranus observations were made (Fig. \ref{fig:beams}). The actual beam pattern differed significantly from a single Gaussian, especially at $450\ \mu\mathrm{m}$ where 55\% of the power was not in the main beam. The actual beam pattern, $B(\theta)$, was approximated by a superposition of Gaussians, i.e.,
\begin{equation}
    B(\theta) = w_1 G_1(\theta, \sigma_1) + w_2 G_2(\theta, \sigma_2) + \dots,
\end{equation}
where the Gaussians were normalised to account for the two dimensional nature of the beams and are given by 
\begin{equation}
    G(\theta, \sigma_i) = \frac{1}{2\pi\sigma^2_i} \exp \left[-\frac{\theta^2}{2\sigma^2_i} \right].
\end{equation}
The relative weights $w_i$ of the Gaussians and their FWHMs ($= 2.35\ \sigma$) are listed in Table \ref{tab:beam}. Three Gaussians were used to characterise the $450\ \mu\textrm{m}$ beam, while two Gaussians were sufficient for the $850\ \mu\textrm{m}$ beam. These approximated beam patterns were used in the interpretation of the observed maps in terms of the modelling as described in Sect. \ref{sec:model}.

\subsection{\label{sec:jcmtline}JCMT Line Observations}
Observations of the \object{W51-IRDC} region were obtained in a variety of molecular lines. With JCMT, maps were obtained in \element[+]{HCO} in both \textit{J}= 3$\rightarrow$2 ($267.6\ \mathrm{GHz}$) and \textit{J}= 4$\rightarrow$3 ($356.7\ \mathrm{GHz}$) transitions on 15 September 2003. These maps were made on a rectangular grid spaced $5''$ in Right Ascension and $10''$ in declination. The maps are shown and discussed in Sect. \ref{sec:line}. Moreover, single pointing observations of \element[+]{HCO} 3$\rightarrow2$ (dated December 5, 2001) and \element{H}\element[+][13]{CO}3$\rightarrow$2 (dated June 15 and July 22, 2004) are available at two positions. Main beam widths are $20\arcsec$ and $14\arcsec$ with efficiencies of $\eta = 0.69$ and $\eta = 0.61$ for the 3$\rightarrow$2 and 4$\rightarrow$3 transitions, respectively, and the velocity resolutions are $0.35\ \mathrm{km\ s^{-1}}$ and $0.53\ \mathrm{km\ s^{-1}}$. Furthermore, similar to the $850\ \mu\mathrm{m}$ continuum case, it was assumed that the beams could be approximated by two Gaussian components where the power of the error beam is given by $1 - \eta$ and its FWHM is derived from the SCUBA's $850\ \mu\textrm{m}$ beam by assuming a scaling proportional to $\lambda$.

All spectra contained low frequency standing waves and a linear baseline subtraction was performed in two windows on opposite sides of the line, covering $15-25\ \mathrm{km\ s^{-1}}$ and $41-51\ \mathrm{km\ s^{-1}}$. The IRDC emission is centered at $v_\textrm{LSR} \simeq 34.0\ \mathrm{km\ s^{-1}}$ with a spatial variation of $\sim 0.5\ \mathrm{km\ s^{-1}}$ over the complex. Using the description of the galactic rotation curve by \citet{1989A&AS...80..149W} this can be translated into a distance of $2.7\ \mathrm{kpc}$. \footnote{The distance assignment from the radial velocity suffers from a basic ambiguity. However, the far distance solution of $8.6\ \mathrm{kpc}$ can be excluded based on the appearance of the IRDC as dark foreground material in front of the bright MIR emission from the inner Galaxy.}

\subsection{Mid-Infrared Extinction\label{sec:mirext}}
\begin{figure}[tp]
    \includegraphics[width=88mm]{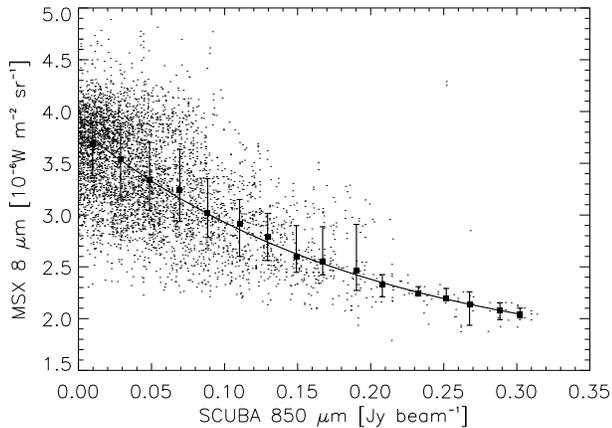}
    \caption{This scatter plot shows the anti-correlation between $8.3\ \mu\mathrm{m}$ emission and $850\ \mu\mathrm{m}$ emission. To model the trend, a median of the MSX emission within bins of $0.02\ \mathrm{Jy\ beam^{-1}}$ was measured. The squares and associated error bars indicate the median values and upper and lower quartiles. The best fitting model to the median data is indicated by the solid line.}
    \label{fig:ext_trend}
\end{figure}
\begin{figure}
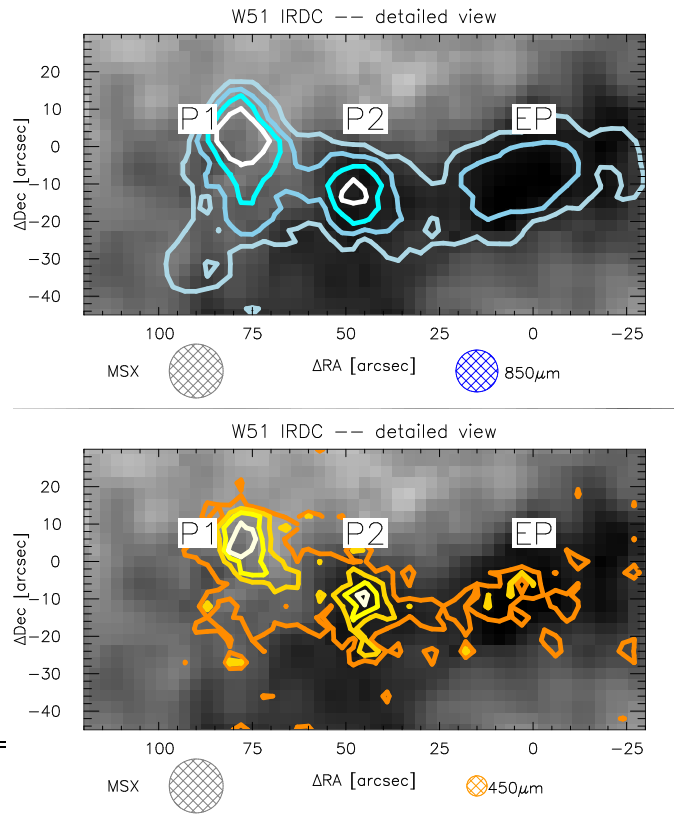

    \resizebox{\hsize}{!}{\includegraphics[bb= 24 90 425 340, clip=true]{2379fg4a.eps}}
    \resizebox{\hsize}{!}{\includegraphics[bb= 24 90 425 340, clip=true]{2379fg4b.eps}}
    \caption{\textit{(top)} Detailed view of the emission peaks, showing P1, P2 and the extinction peak at $850\ \mu\mathrm{m}$. This figure is a close-up of Fig. \ref{fig:W51} with the same contour levels. \textit{(bottom)} $450\ \mu\mathrm{m}$ contour levels drawn at 4,6,8 and 10 times the mean rms-noise with $\sigma_{450} = 0.235\ \mathrm{Jy\ beam^{-1}}$.} 
    \label{fig:W51d}
\end{figure}
\begin{table}
    \begin{center}
    \begin{tabular}{llllll}
    \multicolumn{6}{c}{\sc peak values}\\
    \hline
    \hline
        & \multicolumn{2}{c}{\small Coordinates} &\multicolumn{2}{c}{\small Peak intensities} & $\langle \tau_{8.3\ \mu\textrm{m}} \rangle$ \\[3pt]
        &  R.A. & DEC  & {\small $450\ \mu\mathrm{m}$} & {\small $850\ \mu\mathrm{m}$}  & \\[3pt]
        & \multicolumn{2}{c}{[J2000]} &  \multicolumn{2}{c}{[Jy beam$^{-1}$]} &  \\[3pt]
        & \multicolumn{1}{c}{(1)} & \multicolumn{1}{c}{(2)} & \multicolumn{1}{c}{(3)} & \multicolumn{1}{c}{(4)} & \multicolumn{1}{c}{(5)}\\
    \hline      
    \small
    P1& 19$^h$21$^m$49.9$^s$ & 13\degr49\arcmin34.7\arcsec & 3.08  & 0.531&  $0.4\pm0.4$ \\
    P2& 19$^h$21$^m$47.0$^s$ & 13\degr49\arcmin22.7\arcsec & 2.62  & 0.412 & $1.1\pm0.7$ \\
    EP& 19$^h$21$^m$44.7$^s$ & 13\degr49\arcmin25.7\arcsec & 1.55  & 0.314 & $1.4\pm0.9$ \\
    \hline
    \end{tabular}
    \caption{$450\ \mu\mathrm{m}$ and $850\ \mu\mathrm{m}$ positions, peak intensities and $8.3\ \mu\textrm{m}$ optical depth of P1, P2 and the extinction peak.}
    \label{tab:peaks}
   \end{center}
\end{table}

With images of the region in both the mid-IR and the sub-mm it is possible to estimate the $8.3\ \mu\mathrm{m}$ opacity within the dark cloud complex, together with the background and foreground emission of the Galactic Plane. Following the approach of \citet{2003ApJ...588L..37J}, the measured anti-correlation between $8.3\ \mu\mathrm{m}$ and $850\ \mu\mathrm{m}$ is interpreted as a column density measure in terms of a dust model with constant opacity ratio. Here it is assumed that the foreground and background intensities at $8.3\ \mu\mathrm{m}$,  $I_\mathrm{fg}$ and $I_\mathrm{bg}$, are constant over the map, such that the variation of the measured $8.3\ \mu\mathrm{m}$ intensity can be attributed to the IRDC opacity. When a constant temperature and optical thinness at sub-mm wavelengths are assumed for the IRDC, this opacity, $\tau_{8.3\ \mu\mathrm{m}}$, scales linearly with the $850\ \mu\mathrm{m}$ emission, $\tau_{8.3\ \mu\mathrm{m}} = k I_{850\ \mu\mathrm{m}}$. The measured anti-correlation between $I_{8.3\ \mu\mathrm{m}}$ and $I_{850\ \mu\mathrm{m}}$ can then be used to fit the values of $I_\mathrm{fg}$ and $I_\mathrm{bg}$ with which the $8.3\ \mu\mathrm{m}$ opacity can be computed. Because this fit assumes isothermal and constant dust properties, all positions inside two apertures with radii of, respectively, $27\arcsec$ and $18\arcsec$ centered on the two brightest $850\ \mu\mathrm{m}$ peaks had to be excluded from the derivation of the foreground and background intensities as they clearly belong to sub-mm sources. The resulting scatter plot is shown in \mbox{Fig. \ref{fig:ext_trend}}.

To measure the trend more distinctly, the data were binned in 14 bins of $0.02\ \mathrm{Jy\ beam}^{-1}$.  The solid squares were calculated as the median $8.3\ \mu\mathrm{m}$ emission within each bin. The error estimates are the upper and lower quartiles around this median.

The best fit to the trend gave a foreground emission of \mbox{$I_\mathrm{fg} = (1.6 \pm 0.5) \times 10^{-6}\ \mathrm{W\ m^{-2}\ sr^{-1}}$}, a background emission of \mbox{$I_\mathrm{bg} = (2.3 \pm 0.4) \times 10 ^{-6}\ \mathrm{W\ m^{-2}\ sr^{-1}}$} and a proportionality ratio of $k = 5.1 \pm 2.7$. This is shown in Fig. \ref{fig:ext_trend} as the solid line. 

Assuming a typical dust temperature of $15\ \mathrm{K}$ for the IRDC, the $k$-ratio translates in an mid to far-IR opacity ratio of $\sim870$. This ratio is consistent with the value of $\sim640$ found for another IRDC \citep{2003ApJ...588L..37J} and is close to the value of $\sim500$ predicted for the ice mantle dust grains by \citet{1994AA...291..943O}. In contrast, the diffuse ISM grains of \citet{2001ApJ...554..778L} have a considerably larger ratio of $\sim1700$.

With the fits to the foreground and background emission, a rough estimate for the $8.3\ \mu\mathrm{m}$, MSX-beam averaged, optical depth through the IRDC can be made, i.e.,
\begin{equation}
    \langle \tau_{8.3\ \mu\mathrm{m}} \rangle = - \ln \frac{I - I_\mathrm{fg}}{I_\mathrm{bg}}.
    \label{eq:optdep}
\end{equation}
\begin{figure*}[tp]
\begin{minipage}[l]{1.0\textwidth}
    \includegraphics[width=6cm]{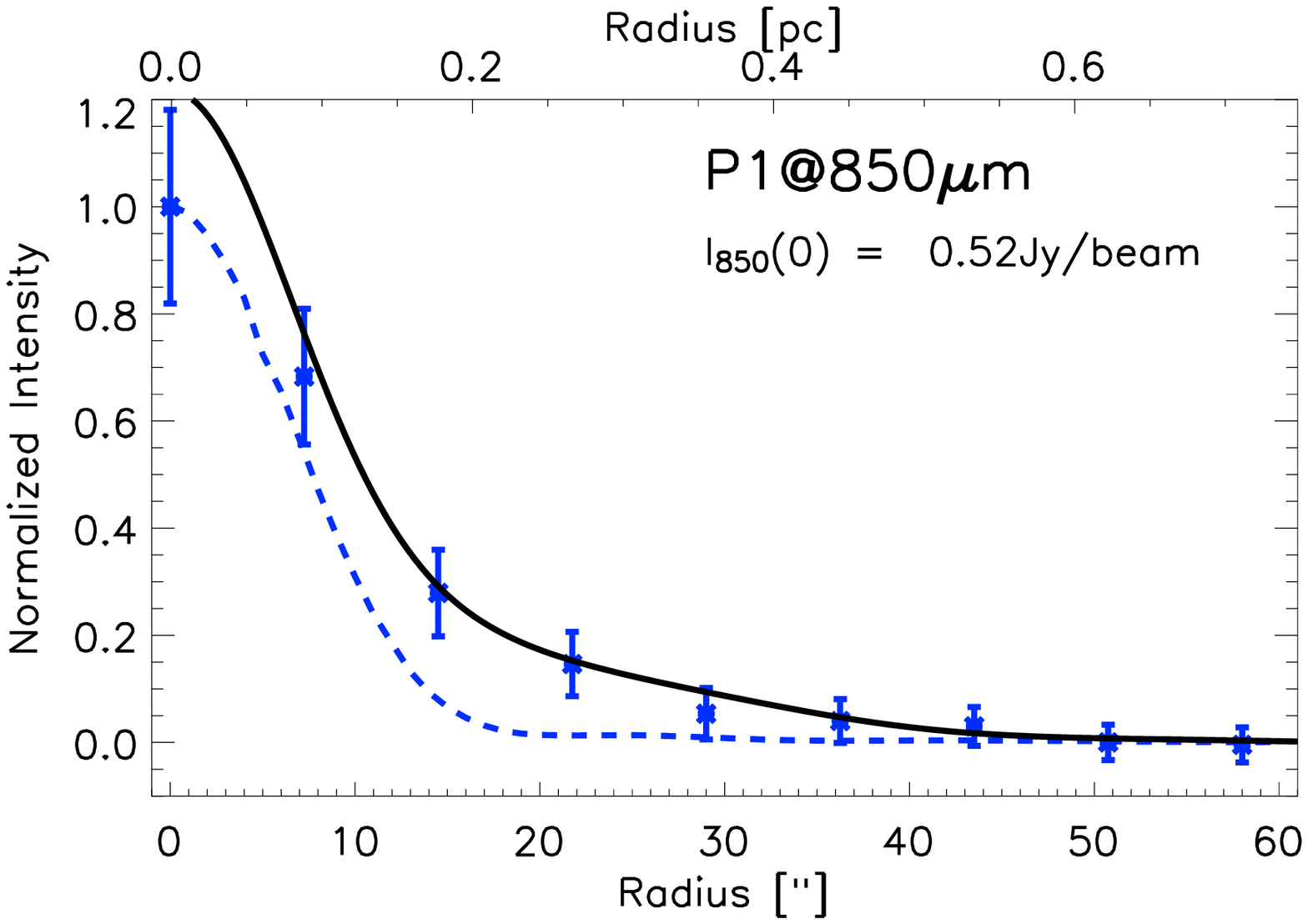}
    \includegraphics[width=6cm]{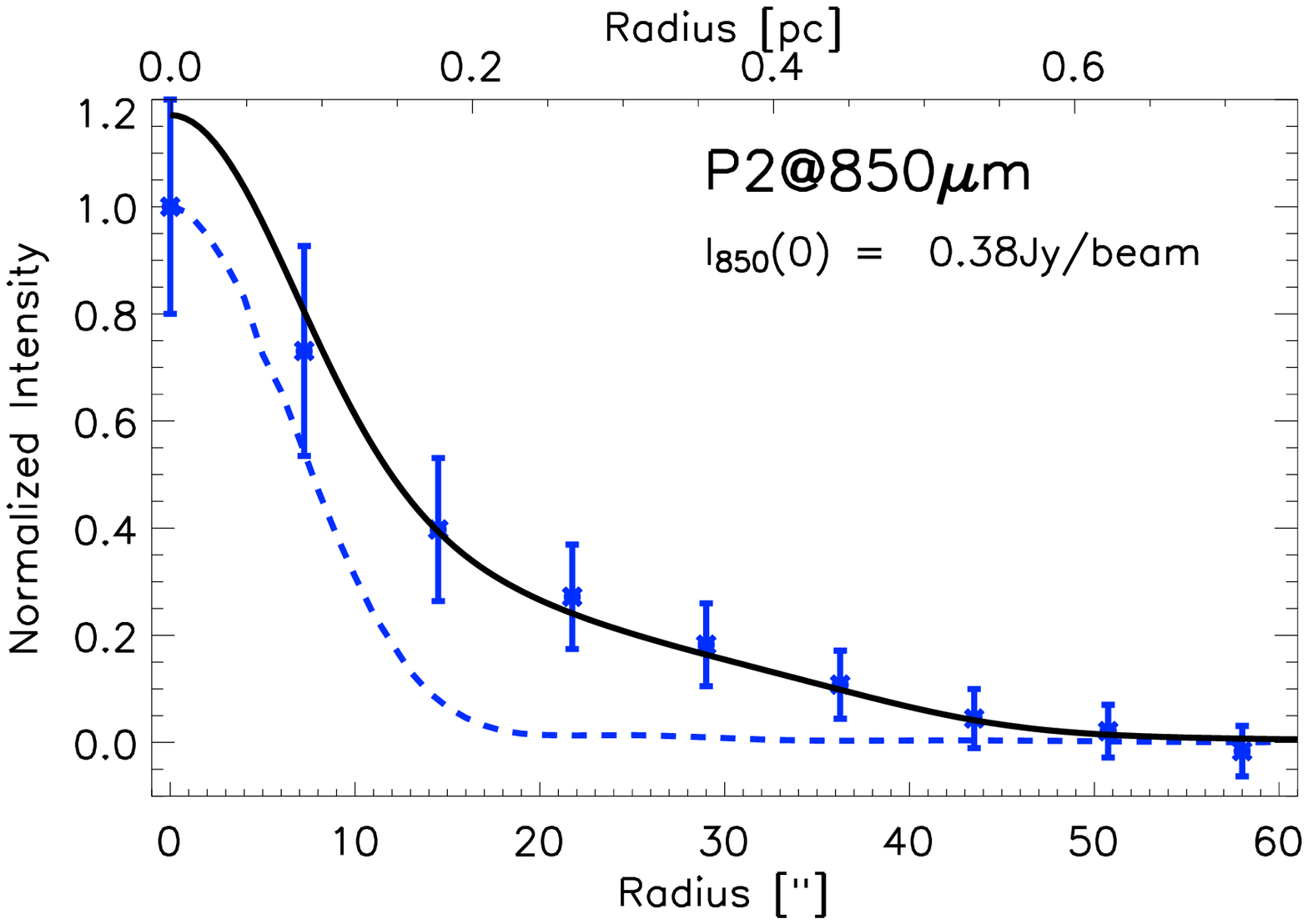}
    \includegraphics[width=6cm]{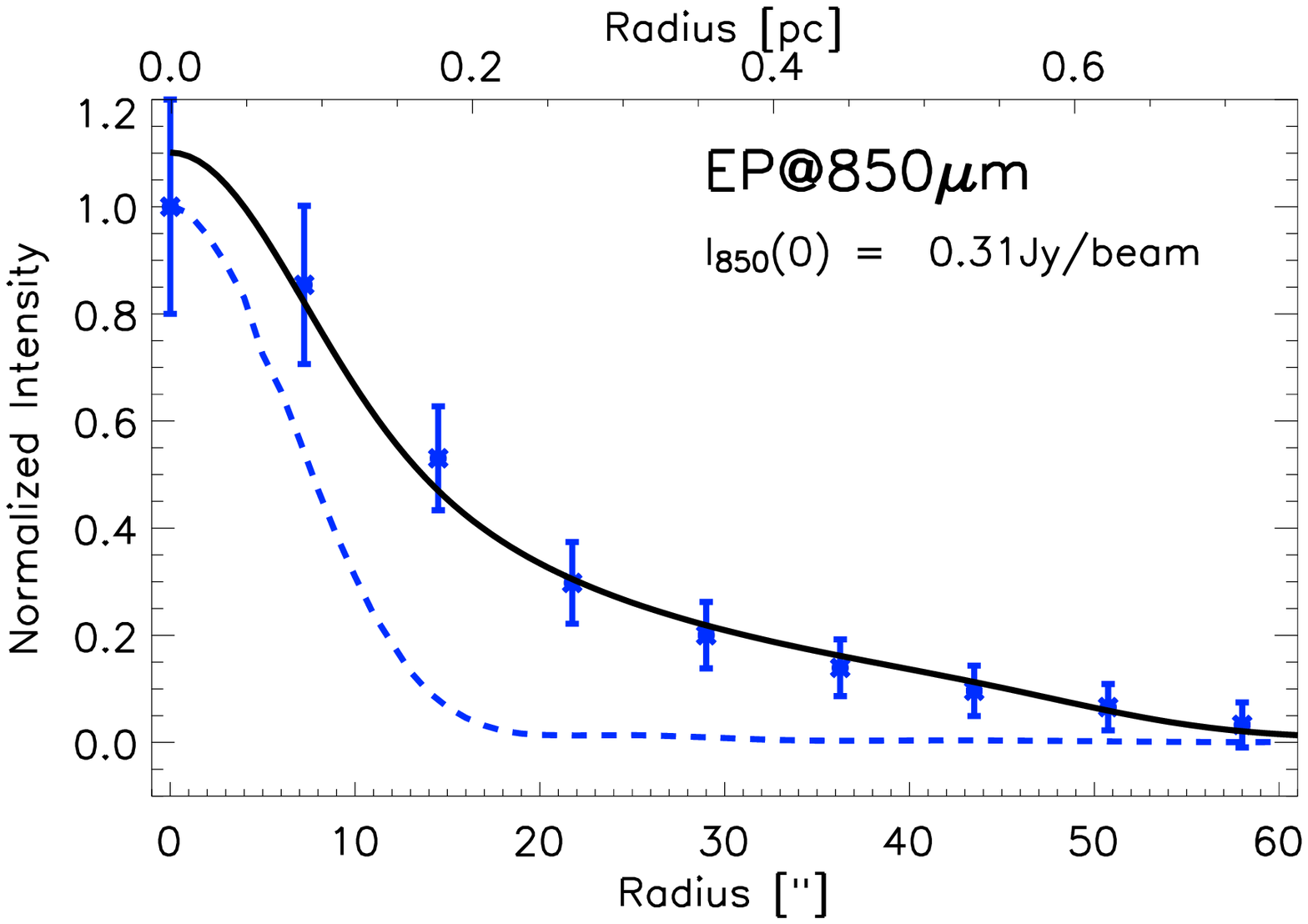}
    \caption{$850\ \mu\mathrm{m}$ radial profiles of P1 (left panel), P2 (centre) and the extinction peak (EP, right). For each panel three features are shown. First the radially averaged data points, binned at every half main beamwidth, where the radial averaging involves directions not contaminated by excess emission from, e.g., neighbouring cores. These points were normalised to their $r = 0''$ values. Second, the dashed line shows the beam profile as obtained from the Uranus observations, which was also normalised to its central value. Finally, the solid line gives the best model fit to the data (see Table \ref{tab:results}).}
    \label{fig:profiles850}
\end{minipage}
\end{figure*}
\begin{figure*}[tp]
\begin{minipage}[l]{1.0\textwidth}
    \includegraphics[width=6cm]{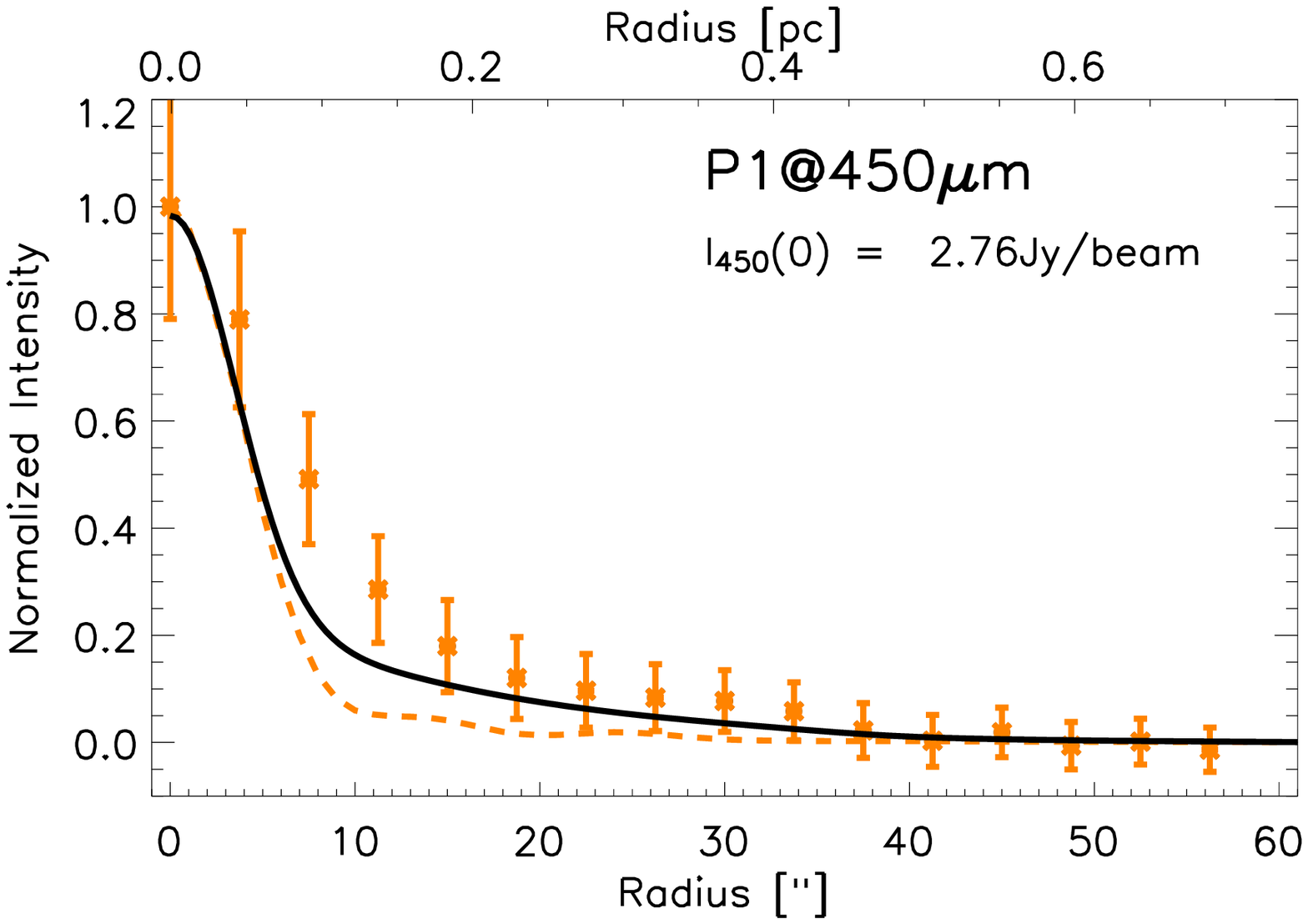}
    \includegraphics[width=6cm]{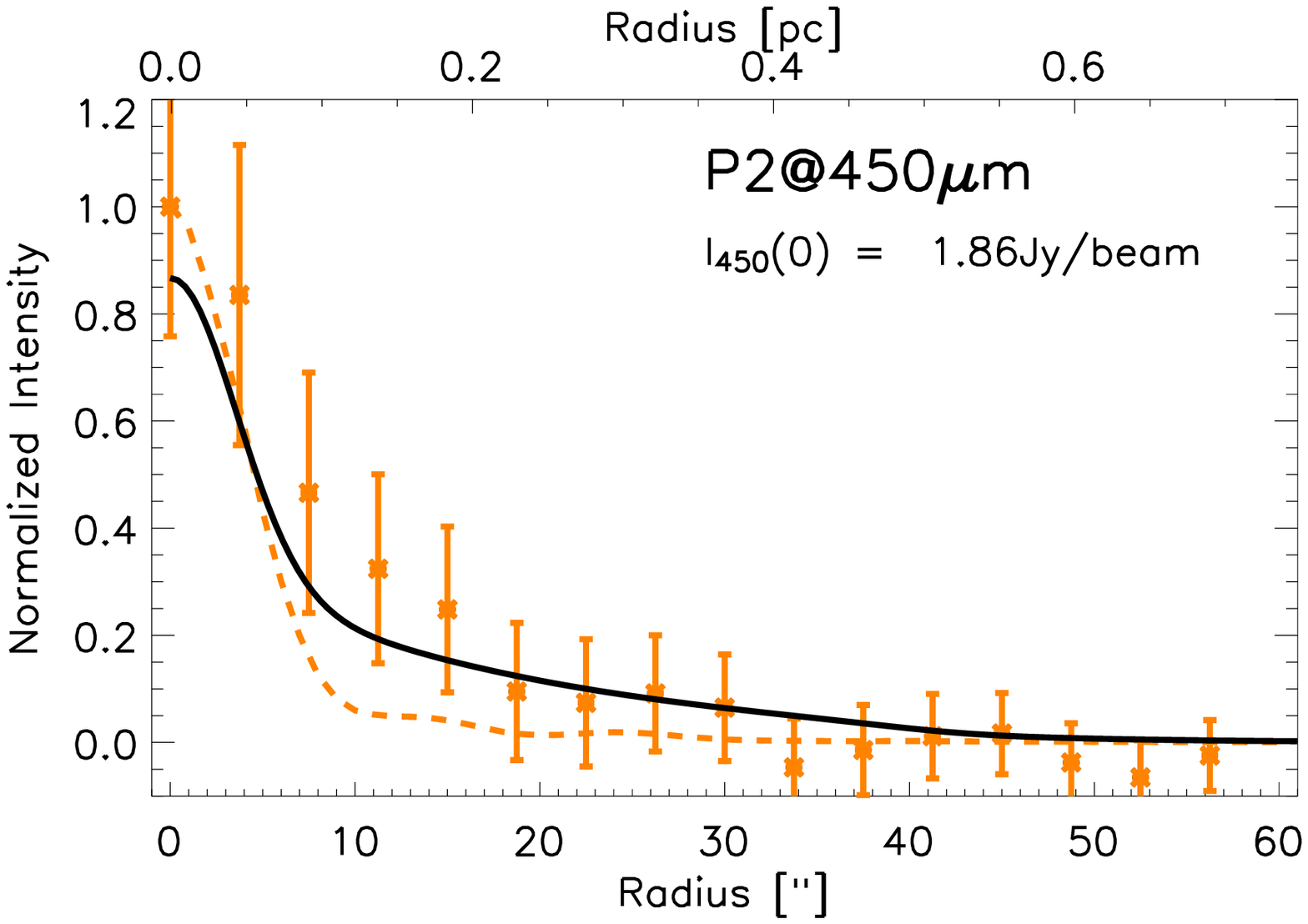}
    \includegraphics[width=6cm]{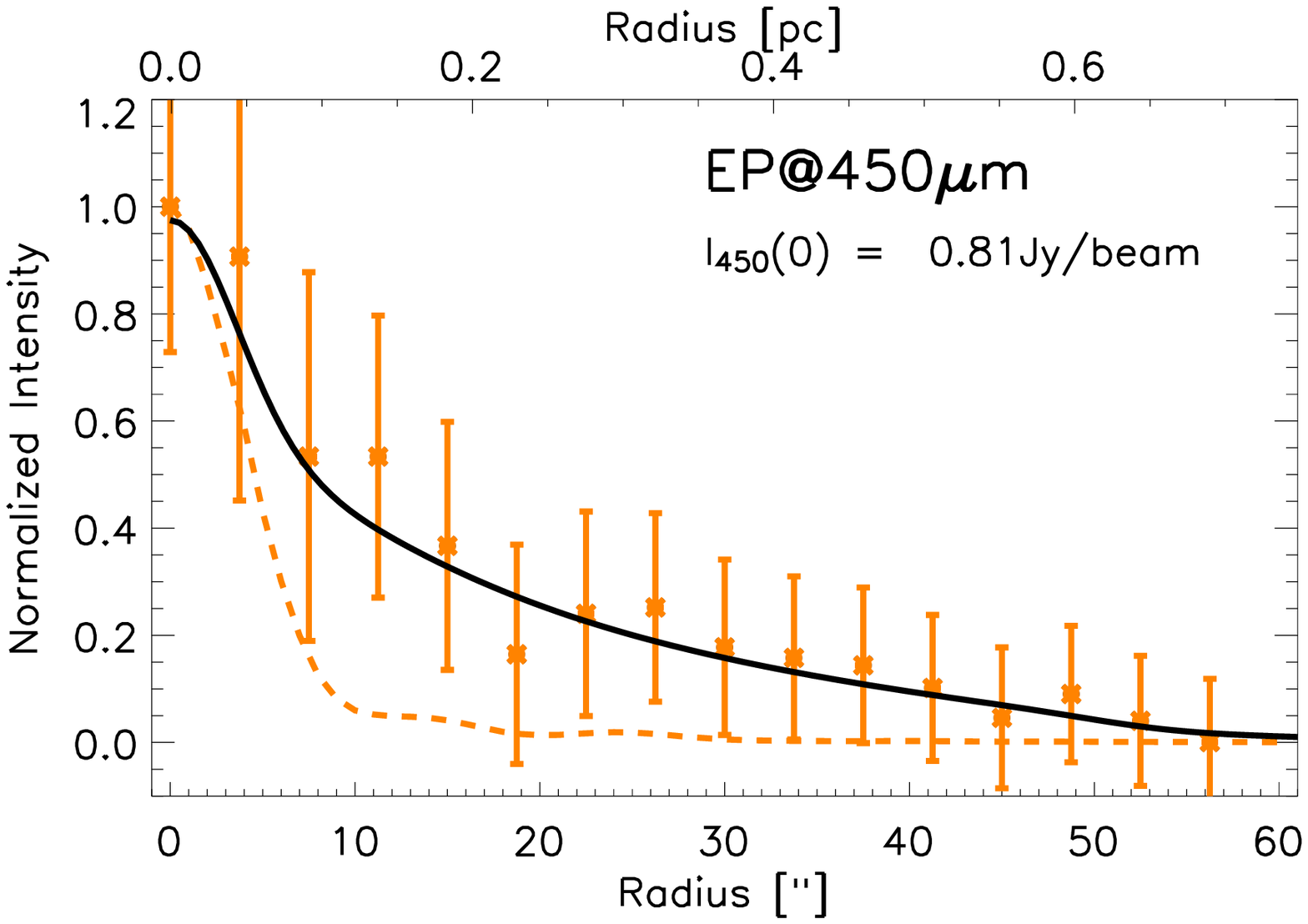}
    \caption{Same as Fig. \ref{fig:profiles850}, but for the $450\ \mu\mathrm{m}$ data.}
    \label{fig:profiles450}
\end{minipage}
\end{figure*}
Where $I$ is the measured intensity from the MSX-map and $I_\mathrm{bg}$ and $I_\mathrm{fg}$ are the estimates from the fit above. The `MSX extinction' therefore anti-correlates with the measured $8.3\ \mu\mathrm{m}$ intensity.
\section{\label{sec:cont}Continuum Modelling}
\subsection{\label{sec:prof}Radial Profiles}

The $450\ \mu\mathrm{m}$ and $850\ \mu\mathrm{m}$ data show three distinct emission peaks (see Fig. \ref{fig:W51d}). The emission peaks are labelled as P1, P2 and EP and the coordinates are listed in Table \ref{tab:peaks}. Table \ref{tab:peaks} also indicates the peak brightness and the estimated extinction as obtained from Eq. (\ref{eq:optdep}). The peak of MSX extinction (position EP) clearly corresponds to an $850\ \mu\textrm{m}$ emission peak. At this position $450\ \mu\textrm{m}$ also peaks, but it is relatively weak. In contrast, the other sub-mm emission peaks are strong at both $850\ \mu\textrm{m}$ and $450\ \mu\textrm{m}$. Note that P1 lies at the edge of the MSX extinction structure, while P2 and EP are well within the complex.

In order to model the emission, one-dimensional radial profiles were constructed from the sub-mm data. Following the approach of \citet{2002ApJS..143..469M}, the sub-mm images are azimuthally averaged over relatively clean regions. The extended IRDC emission was avoided by choosing azimuthal angles for which a radial cut does not pass over other parts of the cloud. 

Figures \ref{fig:profiles850} and \ref{fig:profiles450} show azimuthal averages of the intensity around the three peak positions. Here a zero-level radius of $60\arcsec$ has been introduced such that all profiles go to zero intensity at this radius. In this way low density material at larger scales is ignored, because no reliable information is available for these scales due to the limited chop angle. It is clear that the choice of the zero-level radius introduces systematic uncertainties in the parameters derived from the model. As such the derived radius of the core, $R$, has to be smaller than the zero-level radius and the derived core mass, ${\cal M}(R)$, is limited to the mass well within this zero-level radius. From tests with varying radii we found, however, that the phyiscal structure, i.e., enclosed mass as function of radius, ${\cal M}(r)$, and temperature structure, $T(r)$, are quite insensitive to the applied zero-level radius. The $60\arcsec$ zero-level cut-off applied here is the most natural choice for the boundary between the core and extended material in the IRDC. At these radii the intensity profiles flatten out, indicating  that the emission originates here from diffuse, extended regions rather than the core. One has to keep in mind, however, that the derived total mass of the cores is essentially constrained by this definition of the zero-level radius, as there is no sharp outer boundary of the cores.

\subsection{Core Model\label{sec:model}} 
Each position, P1, P2 and the extinction peak was modelled as a spherical core with a radial density distribution falling off as a power-law. The cores are heated by the interstellar radiation field (ISRF) and a possible internal radiation source. The one dimensional diffusion code CSDUST3 of \citet{1988CoPhC..48..271E} was used to calculate the continuum emission from each of the cores. Together with a density power-law exponent, the code requires a dust grain model that provides the opacities, the optical depth through the core at a specific wavelength, the intensity of the ISRF and the luminosity of the internal source.

CSDUST3 allows for a wide range of adjustable parameters. For reasons of efficiency many of these parameters have been fixed throughout the modelling:
\begin{itemize}
\item Because the data show all three cores as roughly circular, only spherical geometries were considered. Although on a larger scale the cores are part of an elongated extended structure, no attempt was made to incorporate this geometry in the modelling. 
\item The dust model of \citet[model 5; OH5]{1994AA...291..943O} was used. This reflects a high density environment where intermediate ice-mantles might have been formed. 
These opacities are consistent with the extinction analysis of Sect. \ref{sec:mirext}. OH5 opacities are given in units of cm$^{2}$ per gram refractory material and the dust-to-hydrogen ratio of \mbox{$1.5\times 10^{-26}\ \textrm{g\ H-atom}^{-1}$} \citep{1984ApJ...285...89D} was applied to convert the opacity to units of $\textrm{cm}^2$ per hydrogen atom.

\item To avoid infinite densities an inner cavity radius $r_\mathrm{c}$ was introduced. The radius of the cavity was chosen to be \mbox{$r_\mathrm{c} = 0.001\ \textrm{pc} = 206\ \textrm{AU}$} which is small compared to the outer radius of the core ($\sim 1\ \textrm{pc}$). \citet{2003AA...409..589H} concluded for similar modelling that their density distribution and resulting intensity profiles were insensitive to the radius of the inner cavity.
\item The wavelength grid consisted of 60 points, distributed between $\lambda = 1\ \mu\mathrm{m}$ and \mbox{$1\,300\ \mu\mathrm{m}$}. The spatial grid consisted of 100 points and was chosen to have a density of grid points proportional to $r^{-2}$.
\item The model makes use of a `reference frequency' to which the opacity is scaled. $450\ \mu\mathrm{m}$ was chosen as the reference.
\item A description of the ISRF is given by \citet{1983AA...128..212M} and their ISRF at a Galactocentric distance of $6\ \textrm{kpc}$ was used. Since the wavelength grid only starts at $\lambda = 1\ \mu\mathrm{m}$, the UV and part of the near-IR components have been left out. This is justified since the cores lie well embedded within the molecular cloud.
\item In cases where internal heating was included, an effective temperature, $T_\mathrm{eff}$, of this `heating' was fixed at $6\,500\ \textrm{K}$ for all models. A higher value of $T_\mathrm{eff}$ had no effect on the temperature structure, since the high energy photons are quickly re-radiated. The emergent profiles are only affected when the $T_\mathrm{eff}$ of the inner heating becomes very low ($T_\mathrm{eff} \la 30\ \mathrm{K}$).
\end{itemize}

With these specifications four parameters remain free to optimise: $i)$ the exponent of the density distribution $p$, $ii)$ the optical depth at the reference frequency $\tau_\textrm{rod}$, $iii)$ the size of the cloud, i.e., the outer radius, $R$, and $iv)$ the luminosity of the heating source at its centre, $L$. 

The first three combine to give the mass of the core
\begin{equation}
    {\cal M} = 4\pi m_\mathrm{H}\frac{\mu}{2} \frac{p-1}{3-p} \frac{\tau_\mathrm{rod}}{\sigma_\mathrm{rod}} \frac{1 - A_\mathrm{DR}^{p-3}}{ A_\mathrm{DR}^{p-1} - 1} R^2 \qquad (p \neq 1, 3),
    \label{eq:rodtomass}
\end{equation}
where $m_\textrm{H}$ is the mass of one hydrogen atom, $\mu/2 = 1.18$ the mean molecular weight per H-atom, $\sigma_\mathrm{rod} = 1.01 \times 10^{-25}\ \textrm{cm}^2\ \textrm{H-atom}^{-1}$ the opacity at $450\ \mu\mathrm{m}$ and $A_\mathrm{DR} = R/r_\mathrm{c}$ is the dynamic range. Because the mass of the core is as a physical quantity of more interest than the reference optical depth, ${\cal M}$ was chosen as an input parameter and $\tau_\mathrm{rod}$ was adjusted according to \mbox{Eq. (\ref{eq:rodtomass})}.

The output intensities of the model as functions of impact parameter were convolved with the approximated beam pattern from Table \ref{tab:beam}. For each set of parameters a $\chi^2$ goodness-of-fit was calculated 
\begin{equation}
    \chi^2 = \sum_{i (850)} \left( \frac{I_i - M_i}{\sigma_i} \right)^2 + \sum_{i (450)} \left( \frac{I_i - M_i}{\sigma_i} \right)^2 + \left( \frac{M_{100\ \mu\mathrm{m}}}{\sigma_{100\ \mu\mathrm{m}}} \right)^2,
\end{equation}
with $I$ the observed intensity, $M$ the convolved intensity of the model and $\sigma$ the error in the observation. The respective contributions to the $\chi^2$ denote, respectively:
\begin{itemize}
\item The $450\ \mu\mathrm{m}$ and $850\ \mu\mathrm{m}$ profiles, which contribute the data points of the profiles within the cut-off radius of $60''$, separated by the main beams' half width. 
\item The IRAS non-detection at $100\ \mu\mathrm{m}$. $M_{100}$ is the model intensity at $100\ \mu\mathrm{m}$ convolved with the IRAS beam \mbox{(approximated to be spherical with $\textrm{FWHM} = 220\arcsec$)} and $\sigma_{100}$ is the rms-noise of IRAS, which is estimated at \mbox{$\sigma_{100} = 30\ \mathrm{MJy}\ \mathrm{sr}^{-1}$}.
\end{itemize}

\subsection{Results\label{sec:results} and Analysis}

\begin{table}
    \begin{center}
    \begin{tabular}{lllllll}
    \multicolumn{7}{c}{\sc Results Continuum Model} \\
    \hline
    \hline \\[-7pt]
      Core  & $p$                   & ${\cal M}$            & $R\ ^a$   & $L$                   &$\chi^2$   & $\langle \tau_{8.3\mu m} \rangle$     \\[3pt]
            &                       &$[M_{\sun}]$           & $['']$            & [L$_{\sun}$]          &           &                                       \\[3pt]
      (1)   & (2)                   & (3)                   & (4)               & (5)                   &(6)        & (7)                                   \\
\hline \\[-7pt]
    P1      & $2.2^{+0.3}_{-0.2}$      & $91^{+25}_{-21}$      & $37^{+5}_{-7}$ & $330^{+370}_{-180}$   & 12.6      & 0.53                                  \\[3pt]
    P2      & $2.0^{+0.3}_{-0.5}$      & $100^{+30}_{-26}$     & $41^{+6}_{-9}$ & $300^{+1000}_{-230}$  & 7.5       & 0.55                                  \\[3pt]
    EP      & $2.2^{+0.3}_{-0.3}$      & $130^{+25}_{-24}$     & $51^{+7}_{-7}$ & $19^{+110}_{-17}$     & 2.2       & 0.58                                  \\[3pt]
    \hline
    \multicolumn{7}{p{80mm}}{$^a$ Angular Radius. A linear radius of $0.1\ \mathrm{pc}$ corresponds to \mbox{$R = 7.65\arcsec$.}}\\
    \end{tabular}
    \caption{Results of the profiles that have been fitted. In Cols. (2--5) the optimum values of $p, {\cal M}, R \textrm{\ and\ } L$ are given for which $\chi^2$ (Col. 6) is at its minimum. Indicated in the sub- and superscript of the parameters is the range of this parameter for which the $\chi^2$ increases by 1, indicating the $1\sigma$-limits. Finally, the modelled beam-averaged $\tau_{8.3\ \mu\mathrm{m}}$ (Col. 7) is given.}
    \label{tab:results}
  \end{center}
\end{table}
To find the $\chi^2$ minimum in the 4-dimensional parameter space the problem was reformulated as an inverse problem for which solution (or retrieval) techniques are available \citep[e.g.,][]{2000icdi.conf..355B}. An iterative scheme was applied and the absolute minimum of $\chi^2$ was usually found in a few iterations. Moreover, $\chi^2$ contour plots were made in which two parameters were varied. In this way a good visual impression on the $\chi^2$ topology is obtained.

Table \ref{tab:results} gives for every profile the $\chi^2$ minimum and its location in the $(p,{\cal M}, R, L)$-plane. The errors in the parameters give the maximum range for which $\chi^2$ increases by 1, with the remaining three parameters free to find their optimum values, i.e., by minimising $\chi^2$. This $\Delta \chi^2$ behaves also as a $\chi^2$-distribution with one degree of freedom \citep{1992.conf..319}. The $\Delta \chi^2 = 1$ surface gives therefore the $1\sigma$ confidence level of the parameters. Finally, the (MSX beam averaged) model optical depth at $8.3\ \mu\textrm{m}$ is calculated to provide a comparison with the MSX results.

\begin{figure}[tp]
    \includegraphics[width=88mm]{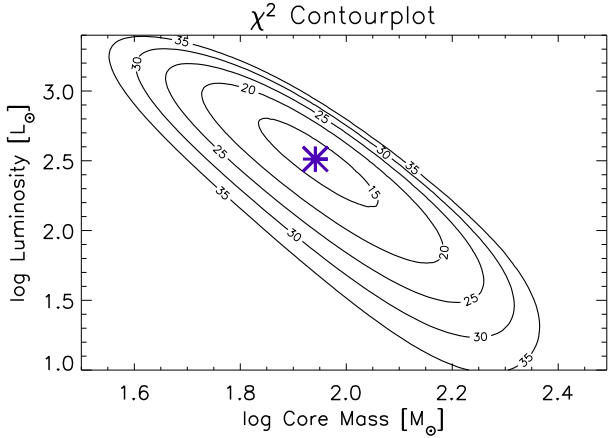}
    \caption{$\chi^2$ contour plot of the P1 model with fixed density exponent ($p = 2.25$) and diameter ($75''$). The solid lines indicate the value of $\chi^2$ in the ${\cal M} - L$ plane. The elongated shape of the $\chi^2$ contours illustrate a basic mass-luminosity degeneracy. The star indicates the $\chi^2$ minimum which is close to the absolute minimum given in Table \ref{tab:results}.}
    \label{fig:chi2map}
\end{figure}
\begin{figure}[tp]
    \includegraphics[width=88mm]{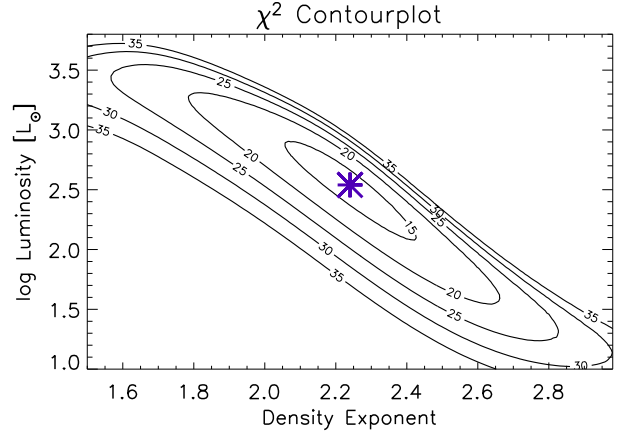}
    \caption{Another contour plot of the P1 model, now at a fixed mass of $90\ \mathrm{M_{\sun}}$ and a diameter of $75''$. The range in luminosity for which good fits ($\chi^2 < 15$) can be obtained spans almost one order of magnitude, while the density exponent can vary between 2.0 and 2.5.}
    \label{fig:chi2map_2}
\end{figure}

Examples of $\chi^2$ contour maps are given in Figs. \ref{fig:chi2map} and \ref{fig:chi2map_2}. In Fig. \ref{fig:chi2map}, $\chi^2$ contours are plotted as functions of mass and luminosity at a fixed size and density exponent, while in Fig. \ref{fig:chi2map_2} density exponent and luminosity are the varied parameters.

We can derive some general conclusions from the $\chi^2$ maps:

\begin{itemize}
\item The IRAS detection limit, estimated at \mbox{$3\sigma \simeq 90\ \mathrm{MJy\ sr}^{-1}$}, corresponds roughly to luminosities of $L = 500 - 1\,000\ \mathrm{L}_{\sun}$, only weakly dependent on mass or density exponent. With luminosities approaching $10^3\ \mathrm{L}_{\sun}$, the contribution of $100\ \mu\mathrm{m}$ to the total $\chi^2$ becomes significant. This sets upper limits to the luminosities of the embedded sources. 
\item For the P1 and P2 models a mass-luminosity degeneracy is observed. Figure \ref{fig:chi2map} gives an example, where both a ${\cal M} = 130\ \mathrm{M}_{\sun}$, $L = 150\ \mathrm{L}_{\sun}$-model and a ${\cal M} = 70\ \mathrm{M}_{\sun}$, $L = 600\ \mathrm{L}_{\sun}$-model yield $\chi^2 = 15$. Core mass and radius are the quantities which are best constrained by the continuum data.
\item Similar to the ${\cal M}-L$ degeneracy a $p-L$ degeneracy can be observed (see Fig. \ref{fig:chi2map_2}). A reduction of the luminosity by a factor of three can be compensated by the increase of the exponent by 0.2.
\item For the extinction peak all $\chi^2$ levels are lower than for the other two peaks. This is, however, only due to the large uncertainties that have been attributed to the $450\ \mu\mathrm{m}$ profile. It is the $850\ \mu\mathrm{m}$ profile that dominates the $\chi^2$-fit. The error bars on the parameters are actually comparable between the cores.
\item For P1 and P2, on the other hand, the main contribution to $\chi^2$ originates from the $450\ \mu\mathrm{m}$ data. 
\end{itemize}
\begin{figure}[tb]
    \centering
    \includegraphics[width=88mm]{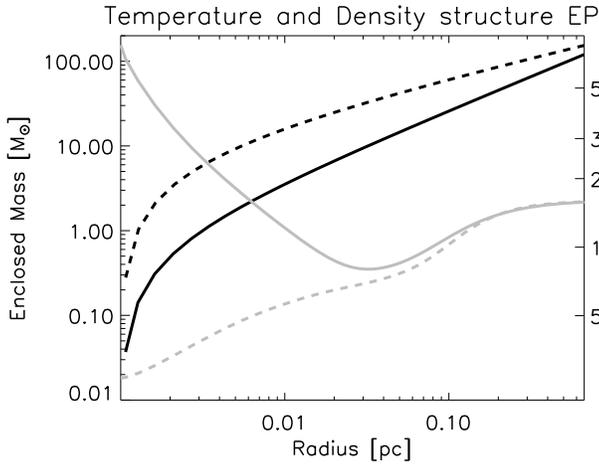}
    \caption{The solid lines give the temperature (grey) and density (black) structure of the minimum $\chi^2$ model of Table \ref{tab:results} for the extinction peak. It shows that temperatures drop below $10\ \mathrm{K}$ within the envelope, but rise toward the boundaries through the heating of the ISRF and the $19\ \mathrm{L_{\sun}}$ source, respectively. The dashed line represents a model without inner heating ($L = 0$). Here, temperatures drop below $5\ \mathrm{K}$ and densities are higher to compensate for the lower temperature. In reality cosmic ray heating would have stabilised the temperature, but the difference would not be noteable from the observations as it is confined to small scales. The $\chi^2$ of this model is 3.9 and its core mass equals ${\cal M} = 170\ \mathrm{M_{\sun}}$.}
    \label{fig:ProfEP}
\end{figure}
\begin{figure}[tb]
    \centering
    \includegraphics[width=88mm]{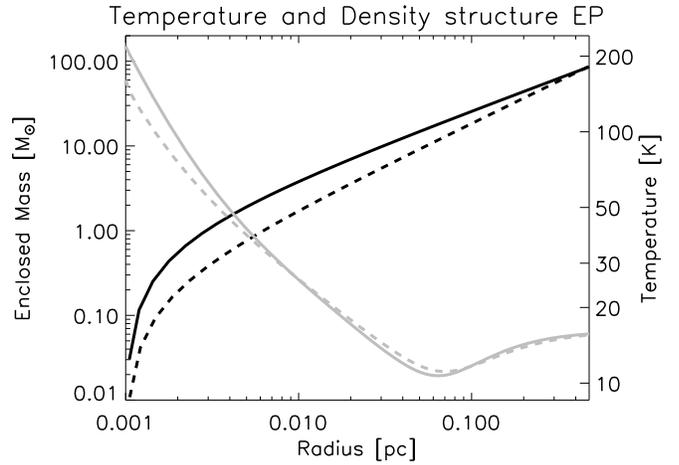}
    \caption{The temperature (grey) and density (black) structure for the minimum $\chi^2$ model of P1 (solid lines) and P2 (dashed lines), respectively. Parameters are given in Table \ref{tab:results}.}
    \label{fig:ProfP1}
\end{figure}
The density and temperature structures that result from the best fit parameters are shown in Figs. \ref{fig:ProfEP} and \ref{fig:ProfP1}. For the extinction peak temperatures have a minimum of about $9\ \mathrm{K}$ within the envelope of the core. At the outer boundaries the temperature is set by the intensity of the ISRF and at the inner boundary by the $19\ \mathrm{L}_{\sun}$ luminous source. The other two peaks (Fig. \ref{fig:ProfP1}), though with higher absolute temperature levels, follow the same qualitative structure. However, since the lower luminosity limit of the extinction peak is only $1.7\ \textrm{L}_{\sun}$, a source-free model is not completely discounted and its density and temperature structure is also given in Fig. \ref{fig:ProfEP}. Here, we have only a marginal indication of an internal heating source from the continuum observations. 

\section{\label{sec:line} \element[+]{HCO} Line Modelling}
\subsection{\label{sec:lines} Lines}
\begin{figure*}
    \centering
    \includegraphics[width=88mm]{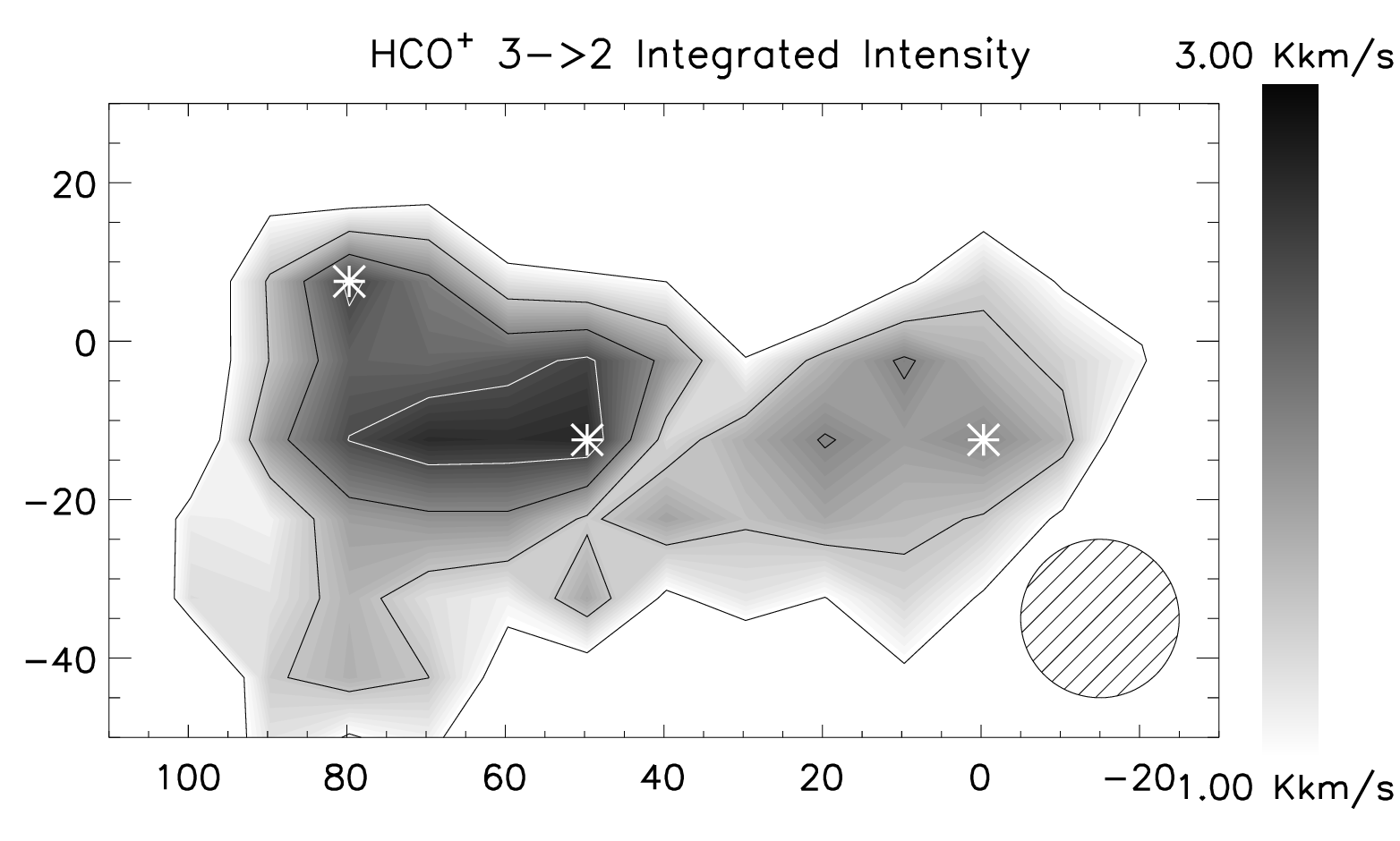}
    \includegraphics[width=88mm]{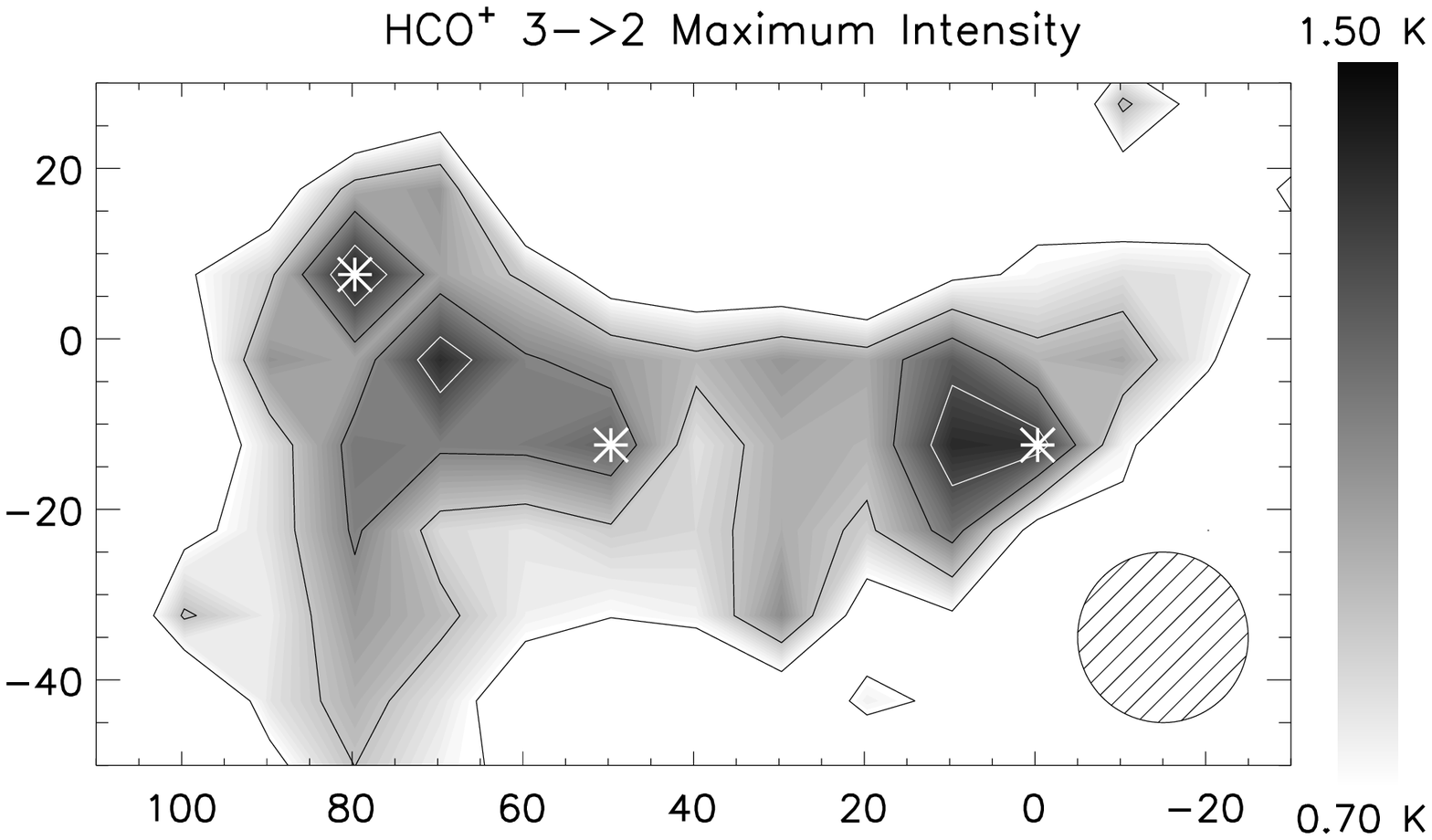}
    \caption{Contour maps of \element[+]{HCO}3$\rightarrow$2. \textit{(left)} Integrated intensity in a window of $v = 34.0\ \pm\ 2\ \mathrm{km\ s^{-1}}$. Contour levels range from $1.0\ \mathrm{K\ km\ s^{-1}}$ to $3.0\ \mathrm{K\ km\ s^{-1}}$ in intervals of $0.5\ \mathrm{K\ km\ s^{-1}}$. The orientation is the same as in the continuum images. The positions of the peaks and the JCMT beamwidth are also indicated. \textit{(right)} The intensity of the highest channel in the $v = 34.0 \pm 2\ \mathrm{km\ s^{-1}}$ window. Here the contours start at $0.70\ \mathrm{K}$ with intervals of $0.2\ \mathrm{K}$.}
    \label{fig:linemaps}
\end{figure*}
The results of the continuum model above provide constraints on the density and temperature structures of the three cores. Modelling of \element[+]{HCO} lines gives information on the velocity structure and, using a turbulent-clumping model, puts constraints on density inhomogeneities. The advantage of using \element[+]{HCO} is that due to its high Einstein A coefficient and the high densities in the cores it radiates easily. Moreover, theoretical studies of \citet{1992MNRAS.255..471R} and observations of the Orion molecular cloud \citep{2003MNRAS.343..259S} show that \element[+]{HCO} is unaffected by the depletion of CO. Recent models of \citet{2004ApJ...617..360L}, on the other hand, claim the opposite. If \element[+]{HCO} is indeed depleted over a wide region of the core, the total abundance will be higher than derived here. However, rather than deriving abundances, the main goal of the line model is to test whether the physical temperature and density structure as derived from the continuum can be confirmed, rejected or further constrained by the line fits. Moreover, we try to extract dynamical information from the lines. A disadvantage of using \element[+]{HCO} is that, contrary to the sub-mm continuum, the lines are optically thick (the models showed typical line-centre optical depths of $\tau \simeq 10$), and the models are especially sensitive to the core's boundary layer.

\begin{table}
    \begin{center}
    \begin{tabular}{lllll}
    \multicolumn{5}{c}{\textsc{\element[+]{HCO} data}}\\
    \hline
    \hline
    Peak    & \multicolumn{2}{c}{Line}                  &  m/p $^a$   &rms-noise \\
            &      &    &                               & [K]      \\
    \hline      
    P1\dots\dots& \element[+]{HCO}\dots                 & 3$\rightarrow$2&m  & 0.21\\
                &                                       & 4$\rightarrow$3&m  & 0.14\\
                & \element{H}\element[+][13]{CO}\dots   & 3$\rightarrow$2&p  & 0.04\\
    P2\dots\dots& \element[+]{HCO}\ldots& 3$\rightarrow$2&m  & 0.27\\
                &                       & 4$\rightarrow$2&m  & 0.14\\
    EP\dots\dots& \element[+]{HCO}\ldots& 3$\rightarrow$2&p  & 0.08\\
                &                       & 4$\rightarrow$3&m  & 0.14\\
                & \element{H}\element[+][13]{CO}\ldots   & 3$\rightarrow$2&p  & 0.03\\
    \hline
    \multicolumn{5}{l}{$^a$ Data as extracted from \textit{m}ap or \textit{p}ointed observations.}
    \end{tabular}
    \caption{An overview of the available lines for every peak. The data have been obtained during several JCMT observing runs (Sect. \ref{sec:jcmtline}).}
    \label{tab:linedata}
    \end{center}
\end{table}

The line maps of the \element[+]{HCO}3$\rightarrow$2 transition are shown in Fig. 11. The emission traces the general structure of the filament. The three cores are best visible in the maximum intensity, but the overall continuum emission is only matched if the information from the maximum intensity map and the integrated intensity map is combined. From the comparison of the two maps it is seen that the lines are narrow at the extinction peak and quite broad at P2. These differences in the lines shapes are exactly reproduced in the following fits of the line profiles.  Perhaps the most notable difference between the two maps is the region east of P2, which is bright in the integrated map. In this region the lines are quite broad. As this broad line region is not associated with a sub-mm core we have not further analysed its properties here, but will be the topic for future studies. It must be emphasised, however, that due to the noisy nature of the spectra, the rather large optical depths and the low resolution, not much information can be obtained from the line maps. Thus no further quantitative conclusions are drawn from the maps apart. 

An overview of the available lines (including pointed observations) is given in Table 4 and the observed lines are shown in Fig. 13. The $S/N$ ratios of the lines reflect the heterogeneity of the observations (see Sect. \ref{sec:jcmtline}). For the extinction peak the \element[+]{HCO}3$\rightarrow$2 pointing has a high $S/N$, while the isotope is barely visible. The \element[+]{HCO} 3$\rightarrow$2 transition of P1 shows a narrow sub-feature near $v \simeq +2.0\ \mathrm{km\ s^{-1}}$, which also appears in the isotope.  At \element[+]{HCO} 4$\rightarrow$3 it is, however, absent. The shape of the sub-feature is suspect as it might be an instrumental artifact or blending with another cold clump. As the nature of the sub-feature is not understood it is assumed at least unrelated to the core and the corresponding channels are masked. However, the assumption that it is merely noise is also tested. In that case the derived parameters change very little relative to the `masked-channels model' with the variation well below the error bars given in Table 5.

The model, explained in more detail below, is structured similar to the continuum model. For a given set of parameters it computes a line spectrum, which, after the convolution with the appropriate beam structure (see Sect. \ref{sec:jcmtline}), is compared to the observed spectral lines. In an iterative fit procedure the model parameters providing the best match to the lines in terms of a $\chi^2$ criterion are obtained. $\chi^2$ is defined as
\begin{equation}
    \chi^2 = \sum_{i,j} \left( \frac{I_{ij} - M_{ij}}{\sigma_{j}} \right)^2,
\end{equation}
where $I_{ij}$ is the intensity in channel $i$ of transition $j$, $M$ the modelled intensity and $\sigma$ the rms-noise in the spectrum. One run of the model includes both \element[+]{HCO} transitions and in the case of P1 and the extinction peak also the \element{H}\element[+][13]{CO} isotope. A value of 70 \citep{1994ARA&A..32..191W} is used for the $\element[+]{HCO} / \element{H}\element[+][13]{CO}$ abundance ratio. For each peak between 28 (EP) and 32 (P1) channels are then simultaneously fitted.

\subsection{\label{sec:simline} Simline}
\begin{figure}
    \centering
    \includegraphics[width=80mm]{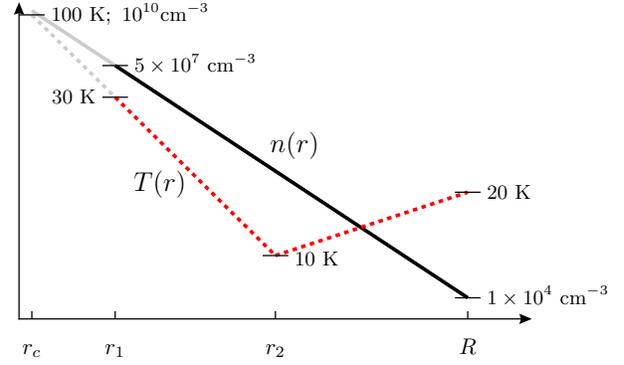}
    \caption{Schematic picture of the density, $n(r)$, and temperature, $T(r)$, structure. For SIMLINE these must be given as power-laws and the temperature structures of Figs. \ref{fig:ProfEP} and \ref{fig:ProfP1} is mimicked using two shells (except for the EP-c model, where one shell suffices). The first shell ranges from $r_1$ to $r_2$. Here $r_1$ rather than $r_c$ is the inner radius and $r_2$ is the position where the temperature is at its minimum. The second shell ranges from $r_2$ to $R$, the core radius. The values given here are those of the clumps and are indicative of P1.}
    \label{fig:physstruc}
\end{figure}
\begin{figure*}[tb]
    \centering
    \includegraphics[width=5.9cm]{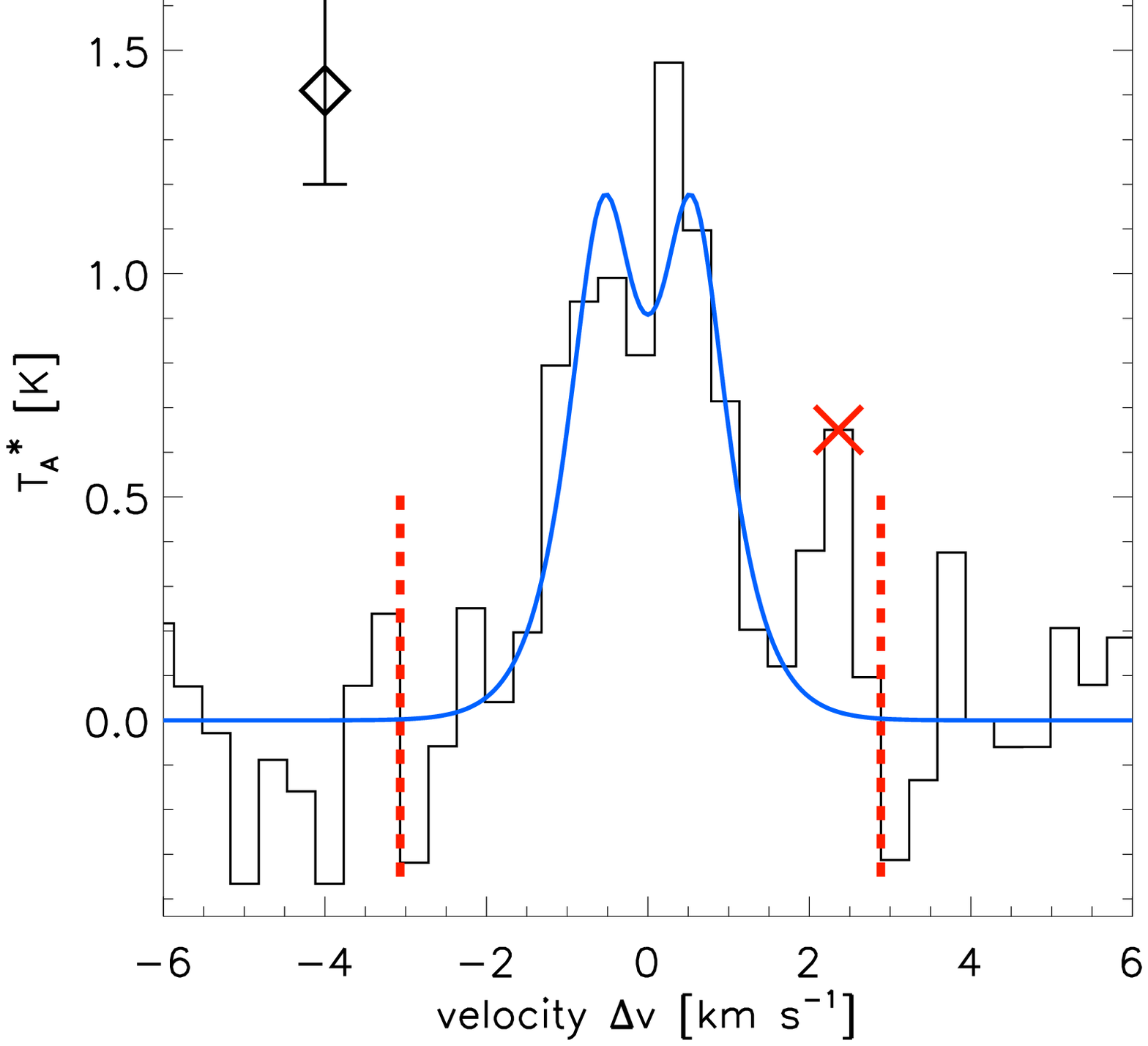}
    \includegraphics[width=5.9cm]{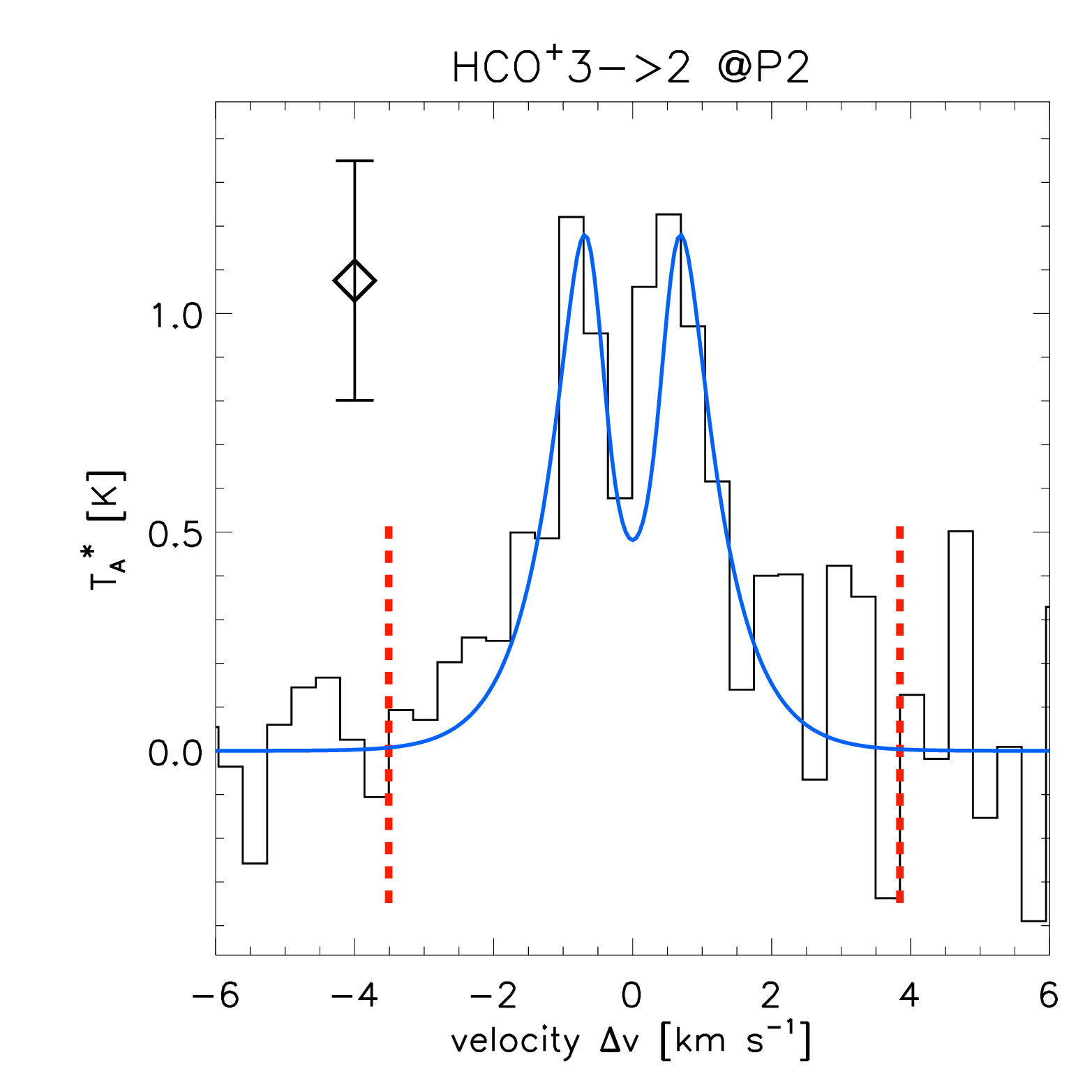}
    \includegraphics[width=5.9cm]{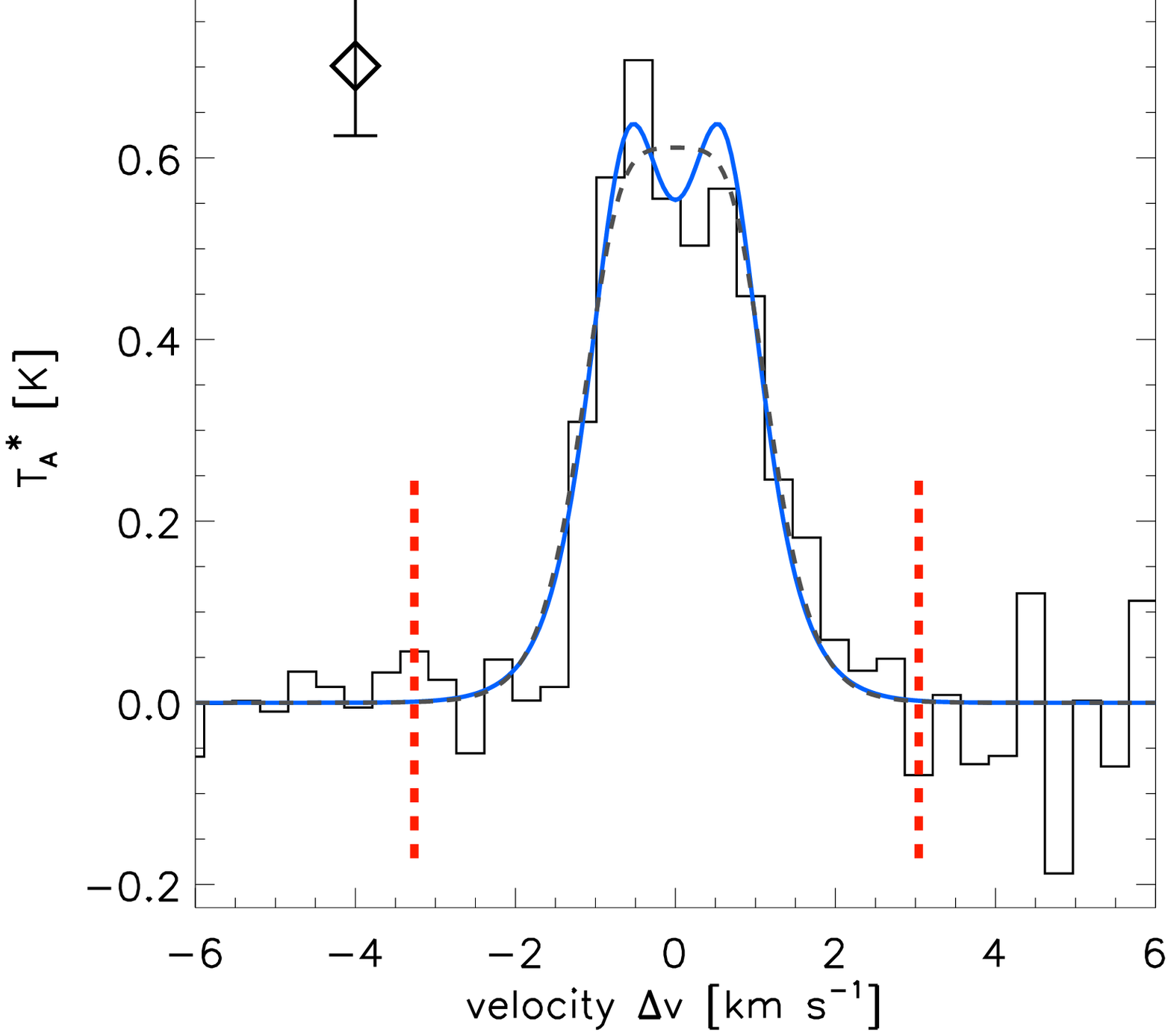}
    \\
    \includegraphics[width=5.9cm]{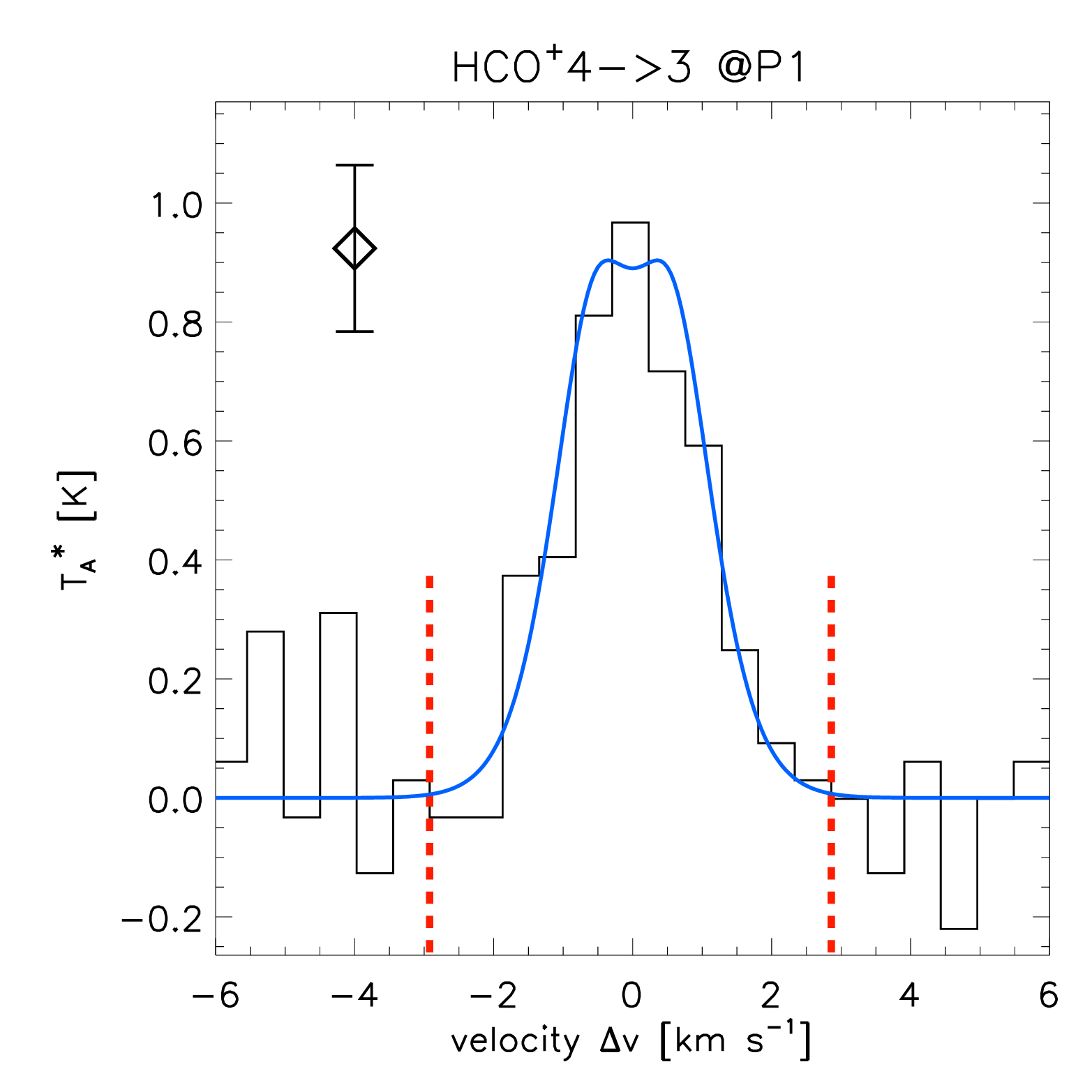}
    \includegraphics[width=5.9cm]{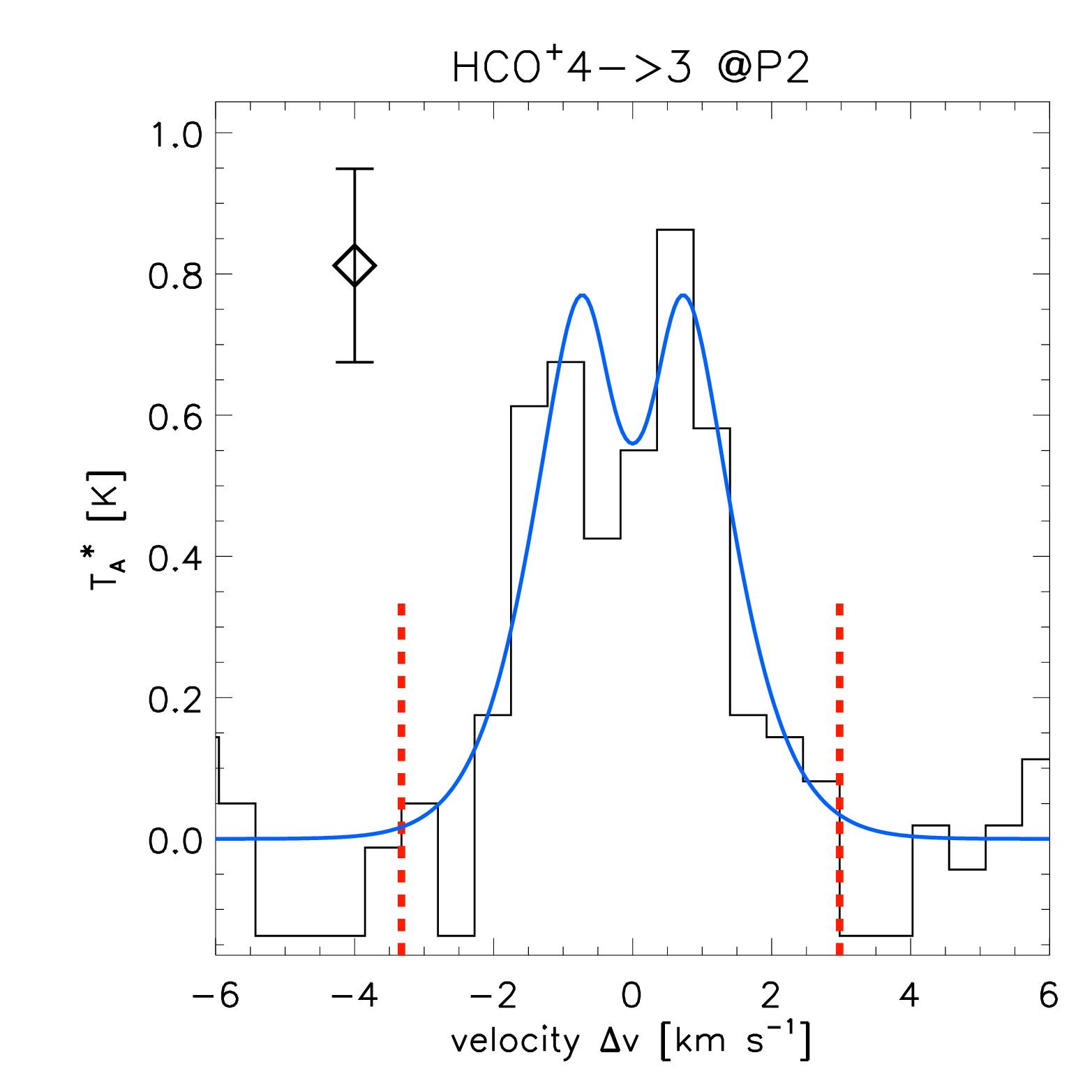}
    \includegraphics[width=5.9cm]{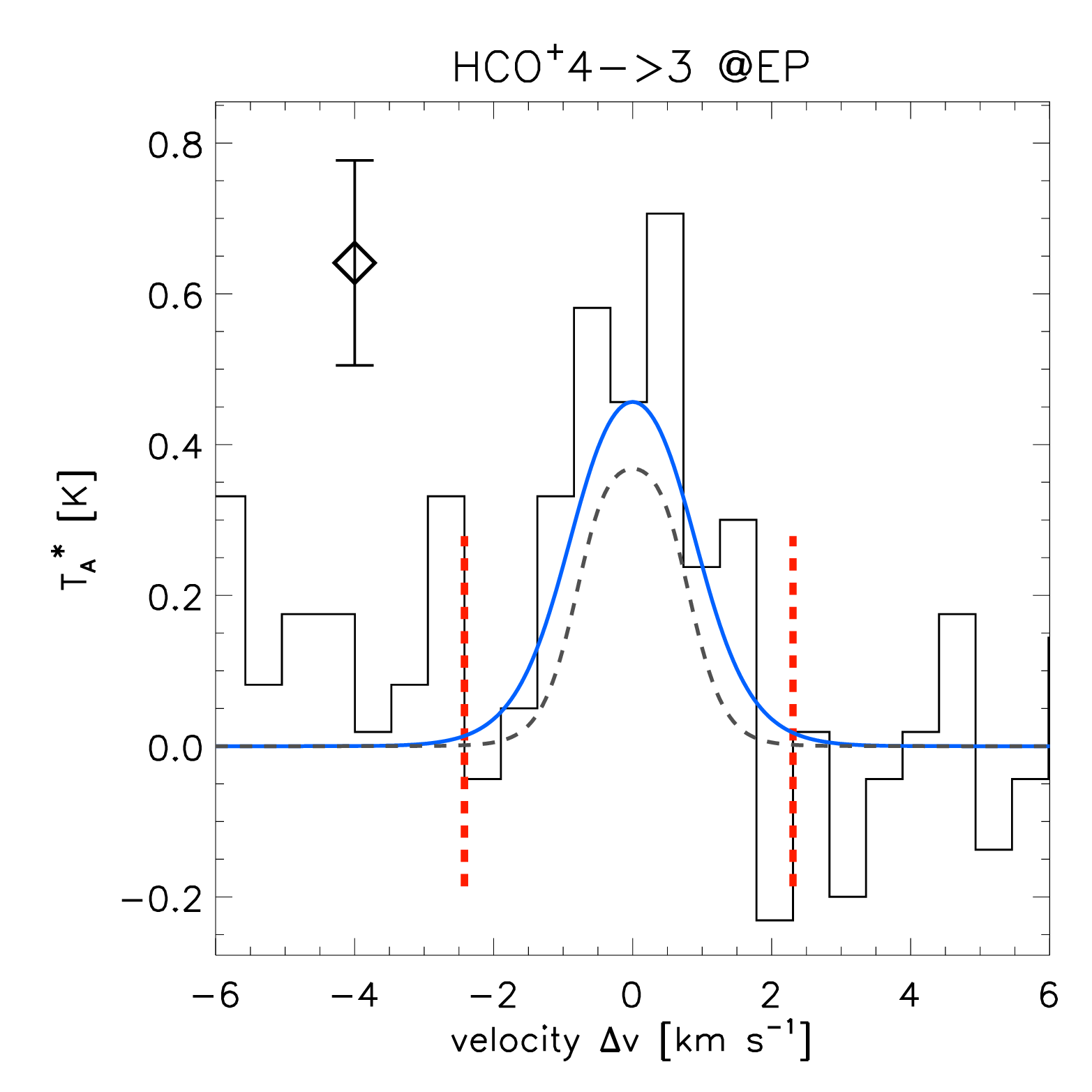}
    \\
    \includegraphics[width=5.9cm]{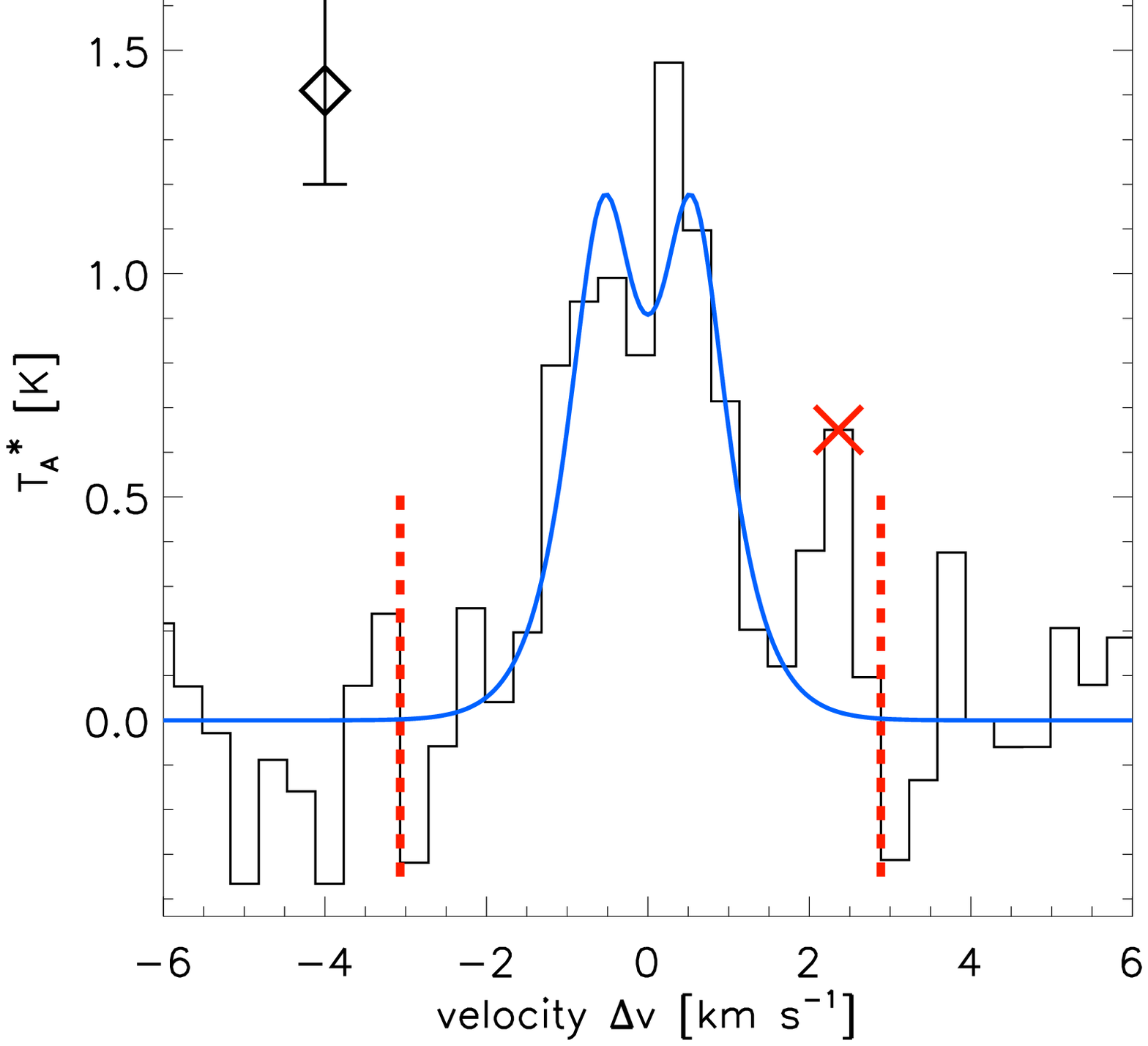}
                   \hspace{5.9cm}
    \includegraphics[width=5.9cm]{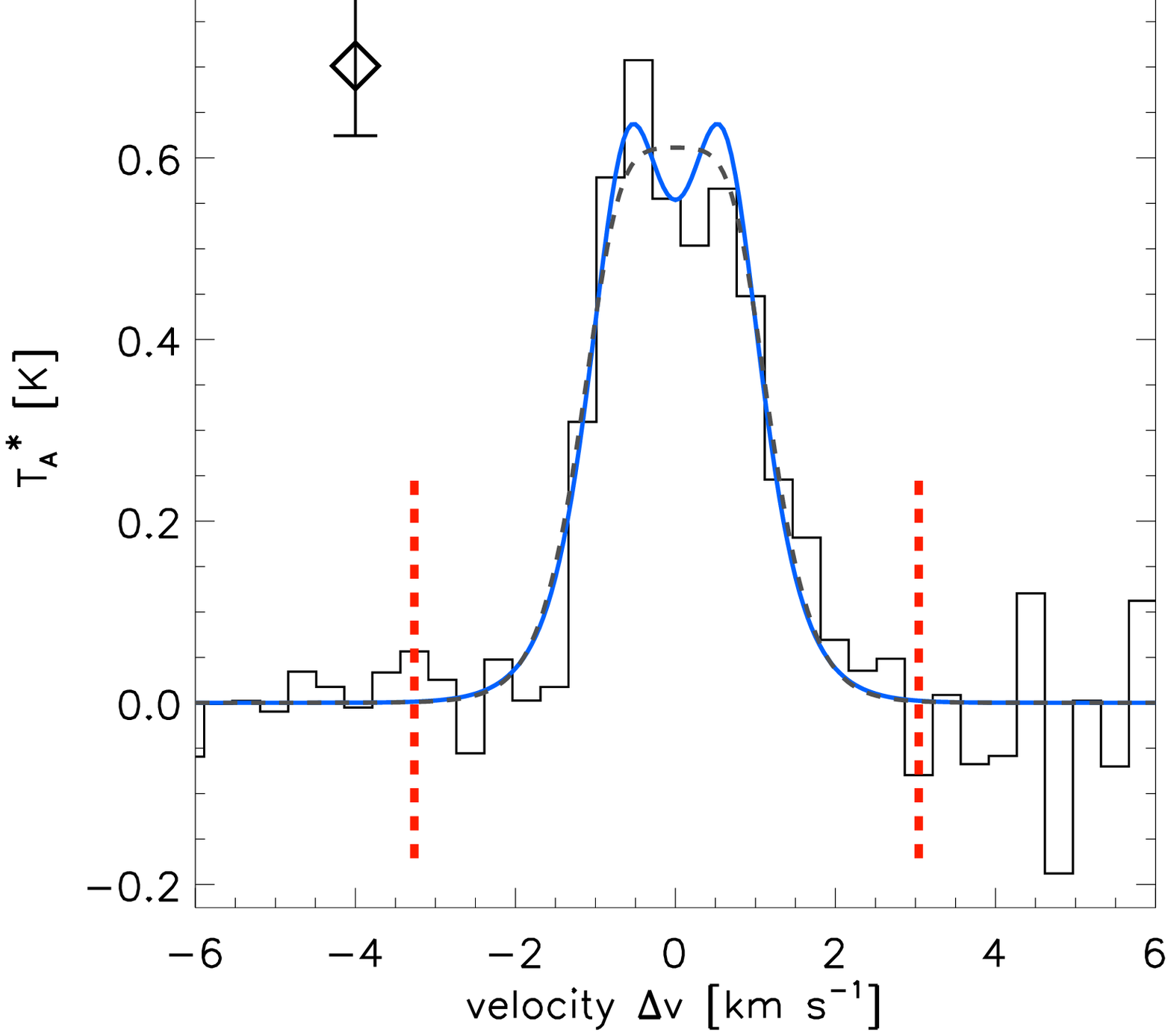}
    \\
    \caption{\element[+]{HCO} observations (histogram) together with the best fits of Table 5 (solid lines). The three rows correspond to the transitions (from top to bottom: \element[+]{HCO}3$\rightarrow$2, \element[+]{HCO}4$\rightarrow$3, H$^{13}$CO$^+$3$\rightarrow$2), while the horizontal direction indicates the model (P1, P2, EP, from left to right). No H$^{13}$CO$^+$ data is available at P2. Furthermore, each panel shows the bounds of the fitted spectral window (dashed vertical lines), the rms-noise level (upper left) and in case of P1 the masked channels (crosses). Velocities are shown relative to $v_\textrm{LSR} = 34.0\ \mathrm{km\ s^{-1}}$. For comparison the EP-plots also show the best fit $L=0$ model with fixed velocity exponent, $v^\mathrm{exp} = 0.5$ (dashed line).}
    \label{fig:spec}
\end{figure*}

The calculation of the model profiles was performed by the spherically symmetric SIMLINE program of \citet{2001A&A...378..608O}. This code is unique since it takes into account effects of turbulent clumping, characterised by a turbulent velocity distribution and the turbulence correlation length $l_\textrm{corr}$. The clumping, which can result in a significant reduction of the effective opacity, is treated in a statistical approach following \citet{1984MNRAS.208...35M}. \citet{2001A&A...378..608O} have shown that this results in a reduction of the strong self-absorption effects from which all (homogeneous) microturbulent models suffer. The code uses a shell structure with parameters that vary as power-laws within the shells and that can be arbitrary from shell to shell. Following the results of the continuum, two shells were used to reflect the temperature structures of Figs. \ref{fig:ProfEP} and \ref{fig:ProfP1} as close as possible (see Fig. 12). 
Since the (clump) densities are high, the gas and dust temperature are expected to be well coupled. At the core radius $R$, however, the gas temperature was slightly augmented to $20\ \mathrm{K}$ to account for PDR-heating of the gas \citep{1989ApJ...338..197S}. Assuming a shielding by the surrounding wide-spread low-density material of $A_V \approx 0.1 - 0.5$ the PDR-heating effects produce only a few degrees difference between gas and dust temperature, so that the assumption of 20~K dust temperature should be accurate within about 1~K. The density structure, determined by $p$, is directly adopted from the continuum (Table \ref{tab:results}). Because it was found that the detailed structure in the centre of the core at radii $r<r_1 = 0.01\ \mathrm{pc}$ had no significant influence on the resulting line profiles, we have omitted this part in the radiative transfer computations. This increased the speed of convergence considerably. 

Four types of models were considered: For each of the peaks the model that corresponds to the parameters of the best fit to the continuum (denoted P1, P2 and EP) and, in the case of the extinction peak, the additional best fit source-free model (denoted EP-c). In this way the line model is self-consistent with the continuum. For each of these models, four input parameters were varied:
\begin{itemize}
\item The turbulent velocity structure (2 parameters: central value and radial exponent). Clouds are often observed to have exponents of 0.5 \citep[e.g.,][]{1999ApJ...524..895M, 2002A&A...390..307O}, while negative exponents indicate that the turbulence is driven from the interior.
\item The \textit{combined} mass. The program simulates a clumpy medium which comprises two components: The (by definition identical) clumps, which fill a volume fraction $f$ and the interclump region, which is treated as void. The density of the clumps is then the average density multiplied with the inverse filling factor $1/f$. Consequently, the combined mass, ${\cal M}^\textrm{com} = {\cal M}/f$, is representative of the conditions inside the clumps. By fitting this parameter for the effective density\footnote{${\cal M}^\mathrm{com}$ is merely a scale function for the density, with the radial gradient fixed by the continuum model.} of the clumps,  and comparing the result with the mass derived from the continuum the filling factor $f$ is obtained.
\item The mass of tracers. In the statistical treatment of the turbulent clumping, it is mathematically equivalent for the radiative transfer problem to treat the clumpy medium as a homogeneous structure with the properties of the clumps, though with a molecular abundance decreased by $f$. Therefore, the output is only sensitive to the combined abundance, $X^\mathrm{com} = Xf$. As the filling factor is already fitted via the combined mass, the molecular abundance of HCO$^+$ can be derived directly by joining the combined mass and combined abundance into the mass of tracers, ${\cal M} X = {\cal M}^\mathrm{com} X^\mathrm{com}$, which we use as the fourth free parameter.
\end{itemize}

\begin{table*}
    \centering
    \begin{tabular}{lllllllll}
    \multicolumn{9}{c}{\sc Results line model} \\
    \hline
    \hline \\[-7pt]
      Model & $v^\textrm{turb}$     & $v^\textrm{exp}$      & ${\cal M}/f$             & $X{\cal M}$     &$\chi^2$   &$\chi^2_\nu$&$f$               &        X     \\[3pt]
           &[$\mathrm{km\ s^{-1}}$] &                       &[$M_{\sun}$]       & [$10^{-7} M_{\sun}$]  &           &       &                       &   [$10^{-10}$]       \\[3pt]
\multicolumn{1}{c}{(1)}&\multicolumn{1}{c}{(2)}&\multicolumn{1}{c}{(3)}&\multicolumn{1}{c}{(4)}&\multicolumn{1}{c}{(5)}&\multicolumn{1}{c}{(6)}&\multicolumn{1}{c}{(7)}&\multicolumn{1}{c}{(8)}&\multicolumn{1}{c}{(9)}\\
    \hline \\[-7pt]
    P1      & $2.6^{+0.8}_{-0.7}$   & $-0.41^{+0.20}_{-0.18}$&$1100^{+170}_{-220}$&$1.6^{+0.3}_{-0.3}  $& 24.7      &0.77   &$0.09^{+0.04}_{-0.03}$ & $17^{+10}_{-6}$  \\[3pt]
    P2      & $3.5^{+0.8}_{-0.5}$   & $-0.50^{+0.09}_{-0.10}$&$530^{+190}_{-180} $&$8.4^{+14}_{-4.2}   $& 24.5      &0.84   &$0.19^{+0.18}_{-0.09}$ & $84^{+210}_{-53}$ \\[3pt]
    EP      & $2.9^{+0.9}_{-0.9}$   & $-0.39^{+0.19}_{-0.15}$&$650^{+120}_{-80}  $&$1.8^{+0.4}_{-0.4}  $& 27.4      &0.98   &$0.20^{+0.07}_{-0.06}$ & $14^{+7}_{-5}$  \\[3pt]
    EP-c    & $3.9^{+1.9}_{-1.4}$   & $-0.50^{+0.23}_{-0.20}$&$910^{+150}_{-130} $&$2.1^{+0.6}_{-0.4}  $& 33.1      &1.18   &$0.19^{+0.07}_{-0.06}$ & $13^{+6}_{-5}$  \\[3pt]
    \hline
    \end{tabular}
    \caption{\label{tab:lineresults}Results of the \element[+]{HCO} line fitting routine. Column (1) shows the four models, the first three correspond to the best fit results of the continuum, while `EP-c' corresponds to the best fitting $L=0$ model for the extinction peak. Cols. (2--5) give the optimum values of the free parameters at which $\chi^2$ (Col. 6) is minimised. The normalised $\chi^2$ value are given in Col. (7). The error limits have the same meaning as in the case of the continuum. Comparing Cols. (4) and (5) with the mass obtained from the continuum model then gives the filling factor, $f$, and the \element[+]{HCO} abundance, $X$.}
\end{table*}
Varying these four parameters is sufficient for the model to fit the data. Therefore, the radial gradient for the velocity structure $v^\textrm{exp}$ is assumed to be the same for both shells, and $X^\textrm{com}$ is assumed to be constant over the whole core. Other parameters that are fixed are the systematic velocities (e.g., the spectra do not show evidence for infall) and the turbulence correlation length, which is fixed at the typical value of $l_\textrm{corr} = 0.01\ \textrm{pc}$. For a more thorough discussion on these radiative transfer parameters see \citet{2001A&A...378..608O}.

\subsection{\label{sec:simres} Results}

Results are presented in Table 5, where the best fit parameters are given. The error limits denote the $1\ \sigma$ boundary of $\chi^2$, calculated in the same way as the continuum. The `new' results are then $v^\textrm{turb}$, $v^\textrm{exp}$, $X$ and $f$, though $X$ and $f$ are not not directly fitted but derived from the fit parameters using the mass values from the continuum fits. The error limits on $f$ and $X$ are calculated by appropriately dividing the upper and lower error limit, e.g., $X^\mathrm{up} = \left( X {\cal M} \right) ^\mathrm{up} / {\cal M}^\mathrm{low}$, where ${\cal M}^\mathrm{low}$ is the lower $1\ \sigma$ mass-limit of the continuum.\\

All fits are satisfactory, as can be seen in Fig. 13, which demonstrates the agreement between fitted and measured profiles. The agreement can be quantified using reduced $\chi^2$ values $\chi^2_\nu=\chi^2/($number of frequency bins -- number of free parameters$)$ (see Table 5). Except for the cold extinction peak model they fall below one and even for this model the value is only slightly above one. \footnote{If the `subfeature' in the P1 lines is not masked, the $\chi^2_\nu$ raises to 1.4. This results only in slight changes of the derived parameters.} Thus the line fit shows again that the model with the internal heating fits the data better, though the cold model still gives reasonable fits. It is remarkable that the errors on the parameters are similar for P1 and the extinction peak in spite of the higher $S/N$ for P1. This can be explained from the equivalent total amount of information for both cores. As the absolute height of the error bars is similar, the relatively weak emission in the isotope for the extinction peak constrains the model in the same way as the stronger emission for P1. In contrast, the error bars of the parameters derived for P2 are considerably larger due to the lack of H$^{13}$CO$^+$ data. When considering the general topology of the $\chi^2$ surface in the considered four-dimensional parameter space in all cases a clear and prominent global minimum was found, similar to the continuum case. By considering many different initial guesses and comparing the results it can be guaranteed that no other local minima were accidentally hit.

No explicit test was made how the uncertainties in the fit results from the continuum modelling affect the outcome of the line fitting. A general estimate for this sensitivity can be obtained, however, from the fit results for the extinction peak.
 Here, we compared a hot and a cold model for the same core. It is found that, although the physical structures of the two models are notably different (see Fig. \ref{fig:ProfEP}), all derived gas-specific parameters are very similar. We therefore anticipate that the errors in the dust model would hardly propagate into larger noticeable errors of the corresponding \element[+]{HCO} models.

Because for the sourceless EP-c model the line fit is not expected to yield negative values for $v^\mathrm{exp}$, we have explicitly tested its influence on the EP-c model by fixing the exponent to 0 and 0.5. This caused the $\chi^2$ of the best fitted model to increase to 36.1 and 39.1, respectively. The increase is due to a slight narrowing of, especially the 4$\rightarrow$3 line. Thus, due to the negative $v^\mathrm{exp}$, the turbulent velocities at the core's centre are large and allow photons to escape through the line-wings, thereby broadening the emergent line-profile. The increase of $\chi^2$ by more than one can be taken as a statistical confirmation that it is valid to fit the $v^\mathrm{exp}$ parameter and that the error limits given in Table 5 are significant. Fig. 13 shows the difference between the two self-consistent models for the extinction peak: The `hot' model where $v^\mathrm{exp}$ is allowed to find its optimum and the `cold', $L=0$, model with $v^\mathrm{exp}$ fixed at 0.5. The $\chi^2$ difference of $\approx 12$, arises mainly due to the $4\rightarrow3$ profile, which the `hot' model fits much better.

\section{\label{sec:discuss}Discussion}

Although the pure $\chi^2_\nu$-values do not exclude the cold model, it can be excluded based on considerations with respect to the dynamical structure of the core. All model fits provide negative values for $v^\textrm{exp}$, i.e., a decrease of turbulent motions outwards of the core. This clearly suggests that the turbulence is driven by star formation activity in the interior of the core. The same source that heats the dust and gas in the centre, is driving turbulent motions which are traced by the HCO$^+$ line profiles. From the combination of the continuum flux profiles and the line ratios, which give both only a relatively weak indication of a heating source in the core, with the line profiles indicating a driving source, it becomes clear that also the extinction peak hosts an active young stellar or protostellar object. The cold model can thus be excluded. For the two sub-mm cores this is much more obvious. All parameters point consistently to luminous sources in both of them.

A significant difference between the fit parameters of the cores is also seen in the molecular abundance. For the P1 and EP models \element[+]{HCO} abundances of around $1.5 \times 10^{-9}$ are found, in good agreement with the results presented by \citet{1997A&AS..124..205H} for warm cores within the \object{W3} complex. Both EP models give about the same value indicating that the derived molecular content does not discriminate between the changes in temperature and density structures of the two models. P2, on the other hand, shows much higher values for the abundance. Even the lower limit of $3.1 \times 10^{-9}$ falls at about twice the average value of the other cores. The reason, nevertheless, for the high fitted abundance can be clearly seen in the line profiles. P2 shows by far the strongest signatures of self-absorption in the profiles, asking for high optical depth in the \element[+]{HCO} line. In contrast one would expect a similar abundance for the three cores, since they resulted from the collapse of the same molecular cloud. A possible explanation for this contradiction is the existence of extended cool foreground material which contributes to the self-absorption profile, but is not detected in the continuum observations due to the limited chop angle which asks for the artificial zero-levelling at large scales. It is also intuitively clear that this effect is strongest for P2 due to its central position in the IRDC. Alternatively, a more complex temperature structure or a systematic change in the internal clumping could also result in an enhanced self-absorption effect. One has to take into account, however, that at least a considerable fraction of the deviation may be due to the large uncertainty in the determination of the abundance for P2. These errors would have been severely constrained if \element{H}\element[+][13]{CO} data, being a good probe for the column density and thus for $X{\cal M}$ due to its low optical depth, were available on P2. Additional observations, both on large scales and with higher resolution, are required to solve this puzzle definitively. 

The values of $0.1-0.2$ found for the filling factors indicate a slightly clumped medium, in agreement with values for other star-forming sites \citep[e.g.,][]{1990ApJ...356..513S,2001A&A...378..608O}. Often even lower values of $f$ can be found \citep[e.g.,][]{1995A&A...294..792H}, though in these models the interclump medium is not treated as void. Apart from the clumpiness no conclusions can be drawn on material at densities $n > 10^6\ \mathrm{cm^{-3}}$, well above the the critical density of the observed lines. The model is, however, sensitive to densities much lower than $n_\mathrm{crit}$. The line profiles provide a strong constraint on the magnitude of self-absorption, thus allowing to derive the structure of the core's outer layers. As a result the model fit infers from the relatively weak self-absorption that the material in the outer layers must be clumped to remain sufficiently excited. For P1, the filling factor is lower by a factor of two and it is tempting to link this low value to the relative transparency of P1 at $8.3\ \mu\textrm{m}$ in the MSX data (see Table \ref{tab:peaks}). In the macroturbulent limit a lowering of $f$ by a factor of two would also decrease the effective optical depth by a factor of two. This is, however, an extreme and the translation of the stronger clumping into a higher $8.3\ \mu\textrm{m}$ transparency is probably less efficient. Thus the clumping alone cannot explain the difference between the optical depths in Tables \ref{tab:peaks} and \ref{tab:results}. Note that the error bars given in Table \ref{tab:peaks} are systematic, i.e., the uncertainties in the background and foreground emission are the same for all peaks, making the average extinction in P2 and EP at least twice the extinction measured for P1, probably even higher. However, it is problematic to compare the findings of Tables \ref{tab:peaks} and \ref{tab:results}. The models deal with the cores only and do not account for the surrounding cloud and the continuum models do not take into account the internal clumping at all.  In summary, the models have clearly shown that the medium must be clumpy on an unresolved scale; Table 5 shows that the line profiles cannot be explained with a homogeneous medium. The stronger turbulent clumping of P1 gives a partial explanation why this core is less prominent in  the MSX data compared to P2 and EP, but it does not explain the full magnitude of the difference.

The key question that triggered this research was whether star formation is taking place within the IRDC-cores. This is obvious for P1 and P2, but for the extinction peak the continuum model allows the sub-mm emission to be explained just by density enhancement. The error bars on the $450\ \mu\textrm{m}$ profile of Fig. \ref{fig:profiles450} are simply too large to break the $p-L$ degeneracy (see Fig. \ref{fig:chi2map_2}): The presence of a possible source can be `hidden' by the model by reverting to a steep density profile. However, the velocity structure, as derived from the \element[+]{HCO} line results, clearly indicates a `hot' model. We have shown, by combining the continuum and line models, that the extinction peak could be one of the earliest stages of star formation observed so far.

The results for the emission peaks, on the other hand, point convincingly to the presence of bright sources, although their luminosities are poorly constrained and could be easily off by a factor of 2--3. The model does not constrain the nature of the sources, though with luminosities of hundreds of solar units, it is hard to think of anything but stellar. Pre main-sequence evolution tracks of \citet{1999ApJ...525..772P} show $300\ \textrm{L}_{\sun}$ luminous sources to correspond to masses in the range of $M \simeq 3-5\ \textrm{M}_{\sun}$, only slightly dependent on the evolutionary stage. This corresponds to intermediate-mass star formation and the spectral type would be late B. However, since star formation in general is not an isolated process, multiple stars might have been formed, which could change this analysis, although it is the highest mass star that produces the bulk of the luminosity.

\section{Conclusions \label{sec:concl}}
The main conclusions of this study are:
\begin{itemize}
\item Through self-consistent modelling of dust emission the presence of warm material embedded in a surrounding envelope is identified for the cores P1 and P2. While the nature of this heating is still unresolved, the energy involved should be at least a few hundreds of solar luminosities. For the extinction peak the presence of a source is not inferred, although the continuum model does prefer it.
\item The molecular line model of \element[+]{HCO} supports the physical structure as obtained from the continuum. The \element[+]{HCO} data reveal the clumpy nature and also provide a strong indication for the existence of an energetic source in the extinction peak. Only from the combination of line and continuum observations can the presence of a star, albeit in a very early evolutionary state, be inferred.
\item The derived luminosities of $\sim 300\ \mathrm{L_{\sun}}$ for P1 and P2 are indicative of intermediate-mass star formation $(M \simeq 3-5\ \mathrm{M_{\sun}})$.
\item The $8.3\ \mu\mathrm{m}$ extinction as derived from the model agrees with the extinction derived from MSX, except for P1, whose lack of $8.3\ \mu\textrm{m}$ extinction in MSX is not in accordance with the model results. The cores' clumpy nature affects their appearance in MSX but the discrepancy is not fully explained. Thus, mid-IR extinction alone cannot identify all cores, especially if star formation is occurring.
\end{itemize}
\begin{acknowledgements}
This research made use of data products from the Midcourse Space Experiment.  Processing of the data was funded by the Ballistic Missile Defense Organization with additional support from NASA Office of Space Science.  This research has also made use of the NASA/ IPAC Infrared Science Archive, which is operated by the Jet Propulsion Laboratory, California Institute of Technology, under contract with the National Aeronautics and Space Administration.
VO was supported by the `Deutsche Forschungsgemeinschaft' through grant SFB 494/A1.
Finally, we thank the anonymous referee whose comments have helped to clarify the paper.
\end{acknowledgements}
\footnotesize
\bibliographystyle{bibtex/aa}
\bibliography{2379publ}
\end{document}